\documentclass[11pt]{article}

\title{\vspace{25mm} \bf{Comments on complete actions for\\ open superstring field theory} \vspace{15mm} }
\author{\Large{Hiroaki Matsunaga}\footnote{matsunaga@fzu.cz} \vspace{5mm}}
\date{${}^{\ast}$Institute of Physics, Academy of Sciences of the Czech Republic, \\ Na Slovance 2, Prague 8, Czech Republic \\ \vspace{2mm}
Yukawa Institute of Theoretical Physics, Kyoto University, \\ Kyoto 606-8502, Japan \vspace{5mm}}

\usepackage[dvips]{graphicx}
\usepackage{amsmath,amscd, amssymb}
\usepackage{cite}
\usepackage{mathrsfs} 
\usepackage{amsfonts}
\usepackage[dvipdfmx, svgnames]{xcolor}
\usepackage{tikz}

\usepackage{upgreek}

\makeatletter
  
  \@addtoreset{equation}{section}
\makeatother

\newcommand{\ld}{ [ \hspace{-0.6mm} [ }
\newcommand{\rd}{ ] \hspace{-0.6mm} ] }
\newcommand{\Ld}{ \big[ \hspace{-1.1mm} \big[ }
\newcommand{\Rd}{ \big] \hspace{-1.1mm} \big] }
\newcommand{\LD}{ \Big[ \hspace{-1.3mm} \Big[ }
\newcommand{\RD}{ \Big] \hspace{-1.3mm} \Big] }

\newcommand{\no}{\nonumber\\}
\newcommand{\niu}[1]{\vspace{7pt}\noindent\underline{{\sf\hspace{4pt}#1\hspace{4pt}}}\vspace{4pt}}

\allowdisplaybreaks[3]

\begin{document}

\maketitle 
{\vspace{-128mm}
\rightline{\tt YITP/15-86}
\vspace{128mm}}

\begin{abstract}
We clarify a Wess-Zumino-Witten-like structure including Ramond fields and propose one systematic way to construct gauge invariant actions: Wess-Zumino-Witten-like complete action $S_{\rm WZW}$. 
We show that Kunitomo-Okawa's action proposed in arXiv:1508.00366 can obtain a topological parameter dependence of Ramond fields and belongs to our WZW-like framework. 
In this framework, once a WZW-like functional $\mathcal{A}_{\eta } = \mathcal{A}_{\eta } [\Psi ]$ of a dynamical string field $\Psi $ is constructed, we obtain one realization of $S_{\rm WZW} [\Psi ]$ parametrized by $\Psi $. 
On the basis of this way, we construct an action $\widetilde{S}$ whose on-shell condition is equivalent to the Ramond equations of motion proposed in arXiv:1506.05774. 
Using these results, we provide the equivalence of two theories: arXiv:1508.00366 and arXiv:1506.05774. 
\end{abstract}

\thispagestyle{empty}
\clearpage

\tableofcontents
\setcounter{page}{1}


\section{Introduction} 

Recently, a field theoretical formulation of superstrings has been moved toward its new phase: An action and equations of motion including the Neveu-Schwarz and Ramond sectors were constructed\cite{Kunitomo:2015usa, Erler:2015lya}. 
With recent developments\cite{Erler:2015rra, Erler:2015uba, Erler:2015uoa, Goto:2015hpa, Konopka:2015tta, Sen:2015hia, Sen:2015uoa, Sen:2015uaa}, we have gradually obtained new and certain understandings of superstring field theories. 
In the work of \cite{Kunitomo:2015usa}, a gauge invariant action including the NS and R sectors was constructed without introducing auxiliary Ramond fields or self-dual constraints. 
They started with the Wess-Zumino-Witten-like action\footnote{Note that the starting NS action is that of $\mathbb{Z}_{2}$-reversed theory and has gauge invariance with $(\delta e^{\Phi })e^{-\Phi } = \eta \Omega - \ld (\eta e^{\Phi }) e^{-\Phi } , \Omega \rd + Q \Lambda $, which is constructed from not an equation of bosonic pure gauge\cite{Berkovits:2004xh} but an equation of $\eta $-constraint. See also \cite{GM2} for these $\mathbb{Z}_{2}$-duals for open superstrings with stubs and closed superstrings. } of the NS Berkovits theory\cite{Berkovits:1995ab} and coupled it to the R string field $\Psi ^{\rm R}$ in the restricted small Hilbert space: $XY \Psi ^{\rm R} = \Psi ^{\rm R}$. 
The dynamical string field is an amalgam of the NS large field $\Phi ^{\rm NS}$ and the R restricted small\footnote{See \cite{Berkovits:2001im, Michishita:2004by, Kunitomo:2013mqa, Kunitomo:2014hba, Kunitomo:2014qla} for other fascinating approaches using Ramond fields in the large Hilbert space. } field $\Psi ^{\rm R}$. 
While the complete action \cite{Kunitomo:2015usa} is given by one extension of WZW-like formulation\cite{Berkovits:1995ab, Berkovits:1998bt, Okawa:2004ii, Berkovits:2004xh, Matsunaga:2013mba, Matsunaga:2014wpa}, the other one, the Ramond equations of motion \cite{Erler:2015lya}, is a natural extension of $A_{\infty }$ formulation for the NS sector\cite{Erler:2013xta, Erler:2014eba}. 
The $A_{\infty }$ formulation provides a systematic regularization procedure of superstring field theory\cite{Witten:1986qs, Wendt:1987zh} in the early days. 
This procedure was extended to the case including Ramond fields and the Ramond equations of motion was constructed by introducing the concept of Ramond number projections in \cite{Erler:2015lya}.

\vspace{2mm}

In this paper, we focus on these two important works \cite{Kunitomo:2015usa} and \cite{Erler:2015lya}, and discuss some interesting properties based on Wess-Zumino-Witten-like point of view. 
Particularly, we investigate the following three topics and obtain some exact results. 
\begin{enumerate}
\item We show that one can add the $t$-dependence of Ramond string fields into the complete action proposed in \cite{Kunitomo:2015usa} and make the $t$-dependence of the action ``topological'', which leads us a natural idea of Wess-Zumino-Witten-like structure including Ramond fields. 
\item We clarify a Wess-Zumino-Witten-like structure including Ramond fields and propose a Wess-Zumino-Witten-like complete action. 
Then, it is proved that one can carry out all computation of our action using the properties of pure-gauge-like and associated fields only. 
The action proposed in \cite{Kunitomo:2015usa} gives one realization of our WZW-like complete action. 
\item On the basis of this WZW-like framework, we construct an action whose equations of motion gives the Ramond equations of motion proposed in \cite{Erler:2015lya}. 
As well as the action proposed in \cite{Kunitomo:2015usa}, this action also gives another realization of our WZW-like complete action: different parameterization of the same WZW-like structure and action. 
\end{enumerate}
These facts provide the equivalence of two (WZW-like) theories \cite{Kunitomo:2015usa} and \cite{Erler:2015lya} on the basis of the same discussion demonstrated in \cite{Erler:2015rra}. 
Then, we can also read the relation giving a field redefinition of NS and R string fields with a partial gauge fixing or a trivial uplift by the same way used in \cite{Erler:2015rra, Erler:2015uba} or \cite{Erler:2015uoa} for the NS sector of open superstrings without stubs. 

\vspace{2mm} 

This paper is organized as follows. 
First, we introduce a $t$-dependence of Ramond string fields and transform the complete action proposed in \cite{Kunitomo:2015usa} into the form which has topological $t$-dependence in section 1.1. 
Then, we clarify a Wess-Zumino-Witten-like structure including Ramond fields. 
In section 2, we propose a Wess-Zumino-Witten-like complete action. 
We show that our WZW-like complete action has so-called topological parameter dependence in section 2.1 and is gauge invariant in 2.2. 
In particular, these properties all can be proved by computations based only on the properties of pure-gauge-like fields and associated fields, which is a key point of our construction. 
(In other words, to obtain the variation of the action, equations of motion, and gauge invariance, one does NOT need explicit form or detailed properties of $F$ acting on $\Psi ^{\rm R}$ and $F \Xi $ satisfying $D^{\rm NS}_{\eta } F \Xi + F\Xi D^{\rm NS}_{\eta } =1$ of \cite{Kunitomo:2015usa}, which would heavily depend on the set up of theory. 
See section 2.4 for a linear map $F$ satisfying $F \eta F^{-1} = D_{\eta }$ and $D_{\eta } F \xi + F\xi D_{\eta } =1$.) 
In section 3, we construct an action reproducing the same equations of motion as that proposed in \cite{Erler:2015lya}. 
For this purpose, it is shown that a Wess-Zumino-Witten-like structure naturally arises from $A_{\infty }$ relations and $\eta $-exactness of the small Hilbert space in section 3.2. 
As well as the action proposed in \cite{Kunitomo:2015usa}, this action also gives another realization of our WZW-like complete action. 
Utilizing these facts, we discuss the equivalence of two theories \cite{Kunitomo:2015usa} and \cite{Erler:2015lya} in section 3.3. 
We end with some conclusions. 
Some proofs are in appendix A. 

\niu{Notation of graded commutators}

In this paper, we write $\ld d_{1} , d_{1} \rd $ for the graded commutator of two operators $d_{1}$ and $d_{2}$, 
\begin{align}
\nonumber 
\ld d_{1} , d_{2} \rd \equiv d _{1} \, d_{2} - (-)^{d_{1} d_{2} } d_{2} \, d_{1} . 
\end{align}
Likewise, we write $\ld A , B \rd _{\ast }$ for the graded commutator of the star product $m_{2}(A, B) \equiv A \ast B$, 
\begin{align} 
\nonumber 
\ld A , B \rd _{\ast } \equiv m _{2} ( A , B ) - (-)^{AB} m_{2} ( B ,A ) . 
\end{align} 
The upper index of $(-)^{A}$ or $(-)^{d}$ denotes the grading of the state$A$ or operator $d$, namely, its ghost number. 
For example, we use $\ld D^{\rm NS}_{\eta } , F \Xi \rd = 1$ or $\ld F \Psi ^{\rm R} , F \Psi ^{\rm R} \rd _{\ast } = 2 m_{2} ( F \Psi ^{\rm R} , F \Psi ^{\rm R} )$. 


\subsection{Complete action and topological $t$-dependence} 

In this section, we use the same notation as \cite{Kunitomo:2015usa}. 
First, we show that one can add a parameter dependence of R string fields into Kunitomo-Okawa's action, and that the resultant action has topological parameter dependence. 
Next, from these computations, we identify a pure-gauge-like (functional) field $A^{\rm R}_{\eta }$ and an associated (functional) field $A^{\rm R}_{d}$ of the Ramond sector. 
We end this section by introducing a Wess-Zumino-Witten-like form of Kunitomo-Okawa's action.

\niu{State space and $XY$-restriction}

First, we introduce the large and small Hilbert spaces. 
The large Hilbert space $\mathcal{H}$ is the state space whose superconformal ghost sector is spanned by $\xi (z)$, $\eta (z)$, and $\phi (z)$. 
We write $\eta $ for the zero mode $\eta _{0}$ and $\mathcal{H}_{S}$ for the kernel of $\eta \equiv \eta _{0}$. 
We call this subspace $\mathcal{H}_{\rm S} \subsetneq \mathcal{H}$ the small Hilbert space, whose superconformal ghost sector is spanned by $\beta (z)$ and $\gamma (z)$. 
Let $P_{\eta }$ is a projector onto the $\eta $-exact states: we can write $\mathcal{H}_{\rm S} = P_{\eta } \mathcal{H}$ because $\eta $-complex is exact in $\mathcal{H}$. 
Following the commutation relation $\eta \xi  = 1 - \xi \eta $ for $\xi = \xi _{0}$ or $\Xi $ of \cite{Kunitomo:2015usa}, we define a projector $P_{\xi } \equiv 1 - P_{\eta }$ onto the not $\eta $-exact states. 
%
%
Note also that for any state $\Phi \in \mathcal{H}$, these projectors act as 
\begin{align} 
\nonumber 
P_{\eta } + P_{\xi } = 1 , \hspace{5mm} P_{\eta }^{2} = P_{\eta }, \hspace{5mm} P_{\xi }^{2} = P_{\xi }, \hspace{5mm} P_{\eta } P_{\xi } = P_{\xi }P_{\eta } = 0 , 
\end{align} 
by definition, and that $P_{\eta }$ acts as the identity operator $1$ on $\Phi \in \mathcal{H}_{\rm S}$ because of $\mathcal{H}_{\rm S} \subset P_{\eta } \mathcal{H}_{\rm S}$. 

Next, we consider the restriction of the state space. 
Let $X$ be a picture-changing operator which is a Grassmann even and picture number $1$ operator defined by $X = \delta ( \beta _{0} ) G_{0} - b_{0} \delta ^{\prime } (\beta _{0} ) $, and let $Y$ be an inverse picture-changing operator which is a Grassmann even and picture number $-1$ operator defined by $Y = c_{0} \delta ^{\prime }(\gamma _{0})$. 
These operator satisfy 
\begin{align} 
\nonumber
X Y X = X , \hspace{5mm} Y X Y = Y , \hspace{5mm}  Q X = X Q , \hspace{5mm} \eta X = X \eta . 
\end{align}
The restricted space is the state space spanned by the states $\Psi \in \mathcal{H}$ satisfying $XY \Psi = \Psi $, on which the operator $XY$ becomes a projector $(XY)^{2} = XY$. 
The restricted small space $\mathcal{H}_{R}$ is the space spanned by the states $\Psi $ satisfying 
\begin{align}
\nonumber 
X Y \, \Psi = \Psi , \hspace{7mm} \eta \, \Psi = 0. 
\end{align} 
We use this restricted small Hilbert space $\mathcal{H}_{R}$ as the state space of the Ramond string field. 
See also \cite{Kazama:1985hd, Terao:1986ex, Yamron:1986nb, Kugo:1988mf}. 
One can quickly check that when $\Psi $ is in $\mathcal{H}_{R}$, $Q \Psi $ is also in $\mathcal{H}_{R}$. 
See (2.25) of \cite{Kunitomo:2015usa}.

\niu{Action}

Let $\Phi ^{\rm NS}$ be a Neveu-Schwarz open string field of the Berkovits theory, which is a Grassmann even and ghost-and-picture number $(0|0)$ state in the large Hilbert space $\mathcal{H}$, and let $\Psi ^{\rm R}$ be a Ramond open string field of \cite{Kunitomo:2015usa}, which is a Grassmann odd and ghost-and-picture number $(1|-\frac{1}{2})$ state in the restricted small Hilbert space $\mathcal{H}_{\rm R}$. 
The kinetic term is given by
\begin{align}
\nonumber 
S_{0} = - \frac{1}{2} \langle \Phi ^{\rm NS} , \, Q \eta \Phi ^{\rm NS} \rangle - \frac{1}{2} \langle  \xi Y \Psi ^{\rm R} , \, Q \Psi ^{\rm R} \rangle , 
\end{align}
where $Q$ is the BRST operator of open superstrings, $\langle A , B \rangle $ is the BPZ inner product of $A,B\in \mathcal{H}$ in the large Hilbert space $\mathcal{H}$. 
As explained in \cite{Kunitomo:2015usa}, we can use both $\xi = \xi _{0}$ and $\xi = \Xi $ for the above $\xi $ in the BPZ inner product. 
Utilizing NS WZW-like functionals $A^{\rm NS}_{\eta } = ( \eta e^{\Phi ^{\rm NS}} ) e^{-\Phi ^{\rm NS} }$, $A^{\rm NS}_{t} = ( \partial _{t} e^{\Phi ^{\rm NS}}) e^{-\Phi ^{\rm NS}}$ and the invertible linear map $F$, the full action is given by 
\begin{align}
\nonumber 
S = - \frac{1}{2} \langle \xi Y \Psi ^{\rm R} , \, Q \Psi ^{\rm R} \rangle - \int_{0}^{1} dt \, \langle  A^{\rm NS}_{t} (t) , Q A^{\rm NS}_{\eta } (t) + m_{2} ( F(t) \Psi ^{\rm R} , F (t) \Psi ^{\rm R} ) \rangle , 
\end{align}
where $m_{2}$ is the Witten's star product $m_{2} (A , B ) \equiv A \ast B$ \cite{Witten:1985cc} and as well as $A^{\rm NS}_{\eta } (t)$ or $A^{\rm NS}_{t} (t)$, $F(t)$ also satisfies $F(t=0) = 0$ and $F (t=1)=F$. 
Note that $F$ has no ghost-and-picture number and satisfies $F\eta F^{-1}=D^{\rm NS}_{\eta }$ and $D^{\rm NS}_{\eta } F \Xi + F \Xi D^{\rm NS}_{\eta } =1$. 
In this paper, we do not need the explicit form of $F$. 
See \cite{Kunitomo:2015usa} or appendix A for the explicit form of $F$. 
(See also section 2.4.) 

\vspace{2mm} 

In this Kunitomo-Okawa's action, only the NS field $\Phi ^{\rm NS}(t)$ has a parameter dependence and the R field $\Phi ^{\rm R}$ does not have it: $\partial _{t} \Phi ^{\rm NS} ( t) \not= 0$ and $\partial _{t} \Psi ^{\rm R} = 0$, where a $t$-parametrized NS field $\Phi ^{\rm NS}(t)$ is a path satisfying $\Phi ^{\rm NS}(t=0) = 0$ and $\Phi ^{\rm NS}(t=1) = \Phi ^{\rm NS}$ on the state space. 
We show that a complete action of open superstring field theory proposed in \cite{Kunitomo:2015usa} can be written as 
\begin{align} 
\label{S KO}
S = 
- \int_{0}^{1} dt \, \Big( \langle \xi Y \partial _{t} \Psi ^{\rm R} (t) , Q F(t) \Psi ^{\rm R} (t) \rangle + \langle A^{\rm NS}_{t} (t) , Q A^{\rm NS}_{\eta } (t) + m_{2} \big( F(t) \Psi ^{\rm R} (t) , F(t) \Psi ^{\rm R} (t) \big) \rangle \Big) 
\end{align} 
using a $t$-parametrized R field $\Psi ^{\rm R} (t)$ satisfying $\Psi ^{\rm R} (t=0) = 0$ and $\Psi ^{\rm R} (t=1) = \Psi ^{\rm R}$. 
Then, we also show that $t$-dependence of (\ref{S KO}) is topological 
\begin{align}
\label{delta S KO}
\delta S = 
- \langle \xi Y \delta \Psi ^{\rm R} , \, Q F \Psi ^{\rm R} \rangle - \langle A^{\rm NS}_{\delta } ,Q A^{\rm NS}_{\eta } + m_{2} ( F \Psi ^{\rm R} , F \Psi ^{\rm R} ) \rangle . 
\end{align}

\niu{Topological $t$-dependence} 

Let $P_{\eta}$ be a projector onto the space of $\eta $-exact states and let $P_{\xi }$ be a projector defined by $P_{\xi } \equiv 1-P_{\eta }$. 
For example, one can use $P_{\eta } = \eta \xi $ and $P_{\xi } = \xi \eta $ where $\xi = \xi _{0}$ or $\Xi$. 
Note that these projectors $P_{\eta }$ and $P_{\xi }$ satisfy $P_{\eta } + P_{\xi } = 1$ on the large Hilbert space $\mathcal{H}$. 
One can check that 
\begin{align}
\label{R Kinetic KO}
\langle \xi Y \partial _{t} \Psi ^{\rm R} (t) , Q F (t) \Psi ^{\rm R} (t) \rangle & = \langle \xi Y \partial _{t} \Psi ^{\rm R} (t) , Q ( P_{\eta } + P_{\xi } ) F (t) \Psi ^{\rm R} (t) \rangle 
\no 
& = \langle \xi Y \partial _{t} \Psi ^{\rm R} (t) , Q \Psi ^{\rm R} (t) \rangle + \langle \xi Y \partial _{t} \Psi ^{\rm R} (t) , \eta X F (t) \Psi ^{\rm R} (t) \rangle 
\no 
& = \frac{\partial }{\partial t} \Big( \frac{1}{2} \langle \xi Y \Psi ^{\rm R} (t) , Q \Psi ^{\rm R} (t) \rangle \Big) -  \langle \partial _{t} \Psi ^{\rm R} (t) , F (t) \Psi ^{\rm R} (t) \rangle  
\end{align} 
using $P_{\eta } ( F \Psi ^{\rm R} ) = \Psi ^{\rm R}$ and $P_{\eta } Q P_{\xi } = \eta X$. 
See appendix A for their BPZ properties. 

\vspace{2mm}

Similarly, as (\ref{R Kinetic KO}), using the $D^{\rm NS}_{\eta }$-exactness of $F \Psi ^{\rm R} = D^{\rm NS}_{\eta } F \Xi \Psi ^{\rm R}$ and the relation 
\begin{align} 
\nonumber 
\partial _{t} \big( F (t) \Psi ^{\rm R} (t) \big) = F (t) \partial _{t} \Psi ^{\rm R} (t) + F (t) \Xi D^{\rm NS}_{\eta } (t) \ld A^{\rm NS}_{t} (t) , F (t) \Psi ^{\rm R} (t) \rd _{\ast } , 
\end{align}
one can also check that 
\begin{align}
\langle A^{\rm NS}_{t} , m_{2} \big( F(t) \Psi ^{\rm R} (t) , F(t) \Psi ^{\rm R} (t) \big) \rangle & = - \frac{1}{2} \langle F (t) \Psi ^{\rm R} (t) , \ld A^{\rm NS}_{t} (t) , F (t) \Psi ^{\rm R} (t) \rd _{\ast } \rangle 
\no 
& = - \frac{1}{2} \langle \Psi ^{\rm R} (t) , F (t) \Xi D^{\rm NS}_{\eta } (t) \ld A^{\rm NS}_{t} (t) , F (t) \Psi ^{\rm R} (t) \rd _{\ast } \rangle 
\no \nonumber  
& = -\frac{1}{2} \langle \Psi ^{\rm R} (t) , \partial _{t} \big( F (t) \Psi ^{\rm R} (t) \big) - F (t) \partial _{t} \Psi ^{\rm R} (t) \rangle  . 
\end{align} 
Furthermore, since $\eta \Psi ^{\rm R} = 0$ and thus $\langle \partial _{t} \Psi ^{\rm R} , F \Psi ^{\rm R} \rangle + \langle F \partial _{t} \Psi ^{\rm R} , \Psi ^{\rm R} \rangle = \langle \partial _{t} \Psi ^{\rm R} , \Psi ^{\rm R} \rangle = 0$, we find 
\begin{align} 
\label{R Interaction KO}
\langle A^{\rm NS}_{t} (t) , m_{2} \big( F(t) \Psi ^{\rm R} (t) , F(t) \Psi ^{\rm R} (t) \big) \rangle & = -\frac{1}{2} \langle \Psi ^{\rm R} (t) , \partial _{t} \big( F (t) \Psi ^{\rm R} (t) \big) \rangle + \frac{1}{2} \langle \partial _{t} \Psi ^{\rm R} (t) , F (t) \Psi ^{\rm R} (t) \rangle   
\no  
& = \frac{\partial }{\partial t} \Big( - \frac{1}{2} \langle \Psi ^{\rm R} (t) , F (t) \Psi ^{\rm R} (t) \rangle \Big) + \langle \partial _{t} \Psi ^{\rm R} (t) , F (t) \Psi ^{\rm R} (t) \rangle . 
\end{align} 
Therefore using $\langle \Psi ^{\rm R} , F \Psi ^{\rm R} \rangle = - \langle \xi Y \Psi ^{\rm R} , \eta X F \Psi ^{\rm R} \rangle $, we obtain 
\begin{align}
\nonumber 
(\ref{R Kinetic KO}) + (\ref{R Interaction KO}) 
=  \frac{1}{2} \Big( \langle \xi Y \Psi ^{\rm R} , \, Q \Psi ^{\rm R} + \eta X F \Psi ^{\rm R} \rangle \Big)
= \frac{1}{2} \langle \xi Y \Psi ^{\rm R} , \, Q F \Psi ^{\rm R} \rangle . 
\end{align}

\vspace{2mm} 

As a result, our $t$-integrated form of the action (\ref{S KO}) becomes 
\begin{align}
\label{Original KO}
S & = - \frac{1}{2} \langle \xi Y \Psi ^{\rm R} , Q F \Psi ^{\rm R} \rangle - \int_{0}^{1} dt \, \langle A^{\rm NS}_{t} (t) , \, Q A^{\rm NS}_{\eta }(t) \rangle  ,
\no 
& = - \frac{1}{2} \langle \xi Y \Psi ^{\rm R} , Q \Psi ^{\rm R} \rangle - \int_{0}^{1} dt \, \langle A^{\rm NS}_{t} (t) , \, Q A^{\rm NS}_{\eta } (t) + m_{2} \big( F(t) \Psi ^{\rm R} , F(t) \Psi ^{\rm R} \big) \rangle . 
\end{align} 
The second line is the original form used in \cite{Kunitomo:2015usa}, but we do not use the second line expression to show that the variation of the action (\ref{S KO}) is given by (\ref{delta S KO}). 
Translation from the second line to the first line is in appendix A. 
Since the variation of the first term of the first line in (\ref{Original KO}) is 
\begin{align}
\nonumber 
\delta \Big( \frac{1}{2} \langle \xi Y \Psi ^{\rm R} , Q F \Psi ^{\rm R} \rangle \Big) =  \langle \xi Y \delta \Psi ^{\rm R} , \, Q F \Psi ^{\rm R} \rangle + \langle A^{\rm NS}_{\delta } , m_{2} ( F \Psi ^{\rm R} , F \Psi ^{\rm R}) \rangle , 
\end{align} 
we obtain (\ref{delta S KO}) and the action $S$ has topological $t$-dependence. 

\niu{Gauge invariance} 

Let $\Omega ^{\rm NS}$ and $\Lambda ^{\rm NS}$ be ghost-and-picture number $( -1 | -1 )$ and $( -1 | 0 )$ NS states of the large Hilbert space $\mathcal{H}$ respectively, and let $\lambda ^{\rm R}$ be a ghost-and-picture number $( 0 | -\frac{1}{2} )$ R state of the restricted small Hilbert space $\mathcal{H}_{R}$: $\eta \lambda ^{\rm R} = 0$ and $XY \lambda ^{\rm R} = \lambda ^{\rm R}$. 
These states naturally appear in gauge transformations of the action. 
The action $S$ has gauge invariances: $\delta _{\Omega ^{\rm NS}}S = 0$ with $\Omega ^{\rm NS}$-gauge transformations 
\begin{align}
A^{\rm NS}_{\delta _{\Omega ^{\rm NS}} } = \eta \Omega ^{\rm NS} - \ld A^{\rm NS}_{\eta } \Omega ^{\rm NS} \rd _{\ast } , \hspace{5mm} \delta _{\Omega ^{\rm NS}} \Psi ^{\rm R} = 0 ,
\end{align}
$\delta _{\Lambda ^{\rm NS}} S = 0$ with $\Lambda ^{\rm NS}$-gauge transformations 
\begin{align}
\nonumber 
A^{\rm NS}_{\delta _{\Lambda ^{\rm NS}} } = Q \Lambda ^{\rm NS} + \Ld F \Psi ^{\rm R} , F \Xi \ld F \Psi ^{\rm R} , \Lambda ^{\rm NS} \rd _{\ast } \Rd _{\ast }, 
\hspace{5mm} 
\delta _{\Lambda ^{\rm NS}} \Psi ^{\rm R} = X \eta F \Xi  D^{\rm NS}_{\eta } \ld F \Psi ^{\rm R} , \Lambda ^{\rm NS} \rd _{\ast },
\end{align}
and $\delta _{\lambda ^{\rm R}} S = 0$ with $\lambda ^{\rm R}$-gauge transformations 
\begin{align}
\nonumber 
A^{\rm NS}_{\delta _{\lambda ^{\rm R}} } = - \Ld F \Psi ^{\rm R} , F \Xi  \lambda ^{\rm R} \Rd _{\ast }, 
\hspace{5mm} \delta _{\lambda ^{\rm R}} \Psi ^{\rm R} = Q \lambda ^{\rm R} -  X \eta F \Xi \lambda ^{\rm R} .
\end{align}

\vspace{2mm} 

Note also that using a new gauge parameter $\Lambda ^{\rm R}$ defined by 
\begin{align}
\Lambda ^{\rm R} \equiv F \Xi \Big( - \lambda ^{\rm R} + \Ld F \Psi ^{\rm R} , \Lambda ^{\rm NS} \Rd _{\ast } \Big) , 
\end{align}
where $\Lambda ^{\rm R}$ belongs to the large Hilbert space and has ghost-and-picture number $(-1|\frac{1}{2})$,
we obtain a simpler expression of $\Lambda $-gauge transformations as follows 
\begin{align}
A^{\rm NS}_{\delta _{\Lambda }} & = Q \Lambda ^{\rm NS} + \Ld A^{\rm R}_{\eta } , \Lambda ^{\rm R} \Rd _{\ast }, 
\\ 
\delta _{\Lambda } \Psi ^{\rm R} & = - P_{\eta } Q \Big( \eta \Lambda ^{\rm R} - \Ld A^{\rm NS}_{\eta } , \Lambda ^{\rm R} \Rd _{\ast } - \Ld  F \Psi ^{\rm R} , \Lambda ^{\rm NS} \Rd _{\ast } \Big) . 
\end{align}

\niu{Ramond pure-gauge-like field $A^{\rm R}_{\eta }$} 

In the rest of this section, we identify a pure-gauge-like field $A^{\rm R}_{\eta }$ and associated fields $A^{\rm R}_{d}$ in the Ramond sector and rewrite the action (\ref{S KO}) into our Wess-Zumino-Witten-like form. 

We write $A^{\rm R}_{\eta }$ for $F \Psi ^{\rm R}$, which is one realization of {\it a Ramond pure-gauge-like field}: 
\begin{align} 
A^{\rm R}_{\eta } & \equiv F \Psi ^{\rm R} . 
\end{align}
By definition, the R pure-gauge-like field $A^{\rm R}_{\eta }$ satisfies $D^{\rm NS}_{\eta } A^{\rm R}_{\eta } = 0$, namely, 
\begin{align} 
\label{R eta} 
\eta A^{\rm R}_{\eta } - \ld A^{\rm NS}_{\eta } , A^{\rm R}_{\eta } \rd _{\ast } = 0 . 
\end{align} 
As we will explain, the $\eta $-exact component $P_{\eta } A^{\rm R}_{\eta }$ appears in the action and its properties is important. 
Since the linear map $F$ satisfies $\xi F = \xi $ for $\xi = \xi _{0}$ or $\Xi$, we quickly find that 
\begin{align} 
\nonumber 
P_{\eta } A^{\rm R}_{\eta } & = \Psi ^{\rm R} , 
\\ \nonumber 
d ( P_{\eta } A^{\rm R}_{\eta } ) & = d \Psi ^{\rm R} , 
\end{align}
where $P_{\eta } = \eta \xi $ is a projector onto the small Hilbert space $\mathcal{H}_{\rm S}$ and $d = Q, \partial _{t} , \delta $ is a derivation operator commuting with $\eta $. 
Then, using $P_{\eta } A^{\rm R}_{\eta }$, we can express {\it the $XY$-projection invariance} of Ramond string fields $X Y \Psi ^{\rm R} = \Psi ^{\rm R}$ by 
\begin{align}
\label{R eta XY}
X Y ( P_{\eta } A^{\rm R}_{\eta } ) & = P_{\eta } A^{\rm R}_{\eta } . 
\end{align}
Note that $P_{\eta } A^{\rm R}_{\eta } \in \mathcal{H}_{R}$. 
Similarly, we introduce {\it a Ramond associated field} $A^{\rm R}_{d}$ by 
\begin{align}
(-)^{d} A^{\rm R}_{d} \equiv F \Xi \Big( d \Psi ^{\rm R} - (-)^{d} \ld A^{\rm NS}_{d} , F \Psi ^{\rm R} \rd _{\ast } - (-)^{d} \eta \ld d , \Xi \rd F \Psi ^{\rm R} \Big) . 
\end{align}
Using properties of $F$, one can check that the R associated field $A^{\rm R}_{d}$ satisfies 
\begin{align} 
\label{R d}
(-)^{d} d A^{\rm R}_{\eta } = \eta A^{\rm R}_{d} - \ld A^{\rm NS}_{\eta } , A^{\rm R}_{d} \rd _{\ast } - \ld A^{\rm R}_{\eta } , A^{\rm NS}_{d} \rd _{\ast } ,  
\end{align}
or equivalently, $(-)^{d} d A^{\rm R}_{\eta } = D^{\rm NS}_{\eta } A^{\rm R}_{d}-\ld A^{\rm R}_{\eta } , A^{\rm NS}_{d} \rd _{\ast }$. 
See appendix A. 
Then, we obtain  
\begin{align}
\nonumber 
\langle \xi Y \partial _{t} \Psi ^{\rm R} , Q F \Psi ^{\rm R} \rangle + \langle A^{\rm NS}_{t} , m_{2} ( F \Psi ^{\rm R} , F \Psi ^{\rm R} ) \rangle = \langle \xi Y \partial _{t} ( P_{\eta } A^{\rm R}_{\eta } ) , Q A^{\rm R}_{\eta } \rangle + \langle A^{\rm NS}_{t} , m_{2} ( A^{\rm R}_{\eta } , A^{\rm R}_{\eta } ) \rangle . 
\end{align}

Utilizing these expressions, the action becomes 
\begin{align}
S [\Phi ^{\rm NS} , \Psi ^{\rm R} ] = - \int_{0}^{1} dt \Big( \langle \xi Y \partial _{t} ( P_{\eta } A^{\rm R}_{\eta } ) , \, Q A^{\rm R}_{\eta } \rangle + \langle A^{\rm NS}_{t} , \, Q A^{\rm NS}_{\eta } + m_{2} ( A^{\rm R}_{\eta } , A^{\rm R}_{\eta } ) \rangle \Big) . 
\end{align}
Note also that gauge transformation parametrized by $\Omega = \Omega ^{\rm NS} + \Omega ^{\rm R}$ is given by 
\begin{align}
\nonumber 
A^{\rm NS}_{\delta _{\Omega } }  = \eta \Omega ^{\rm NS} - \Ld A^{\rm NS}_{\eta } , \Omega ^{\rm NS} \Rd _{\ast } , 
\hspace{5mm}  
\delta _{\Omega } ( P_{\eta } A^{\rm R}_{\eta } ) = 0 . 
\end{align}
and gauge transformations parametrized by $\Lambda = \Lambda ^{\rm NS} + \Lambda ^{\rm R}$ is given by 
\begin{align}
\nonumber 
A^{\rm NS}_{\delta _{\Lambda }} & = Q \Lambda ^{\rm NS} + \Ld A^{\rm R}_{\eta } , \Lambda ^{\rm R} \Rd _{\ast } , 
\\ \nonumber 
\delta _{\Lambda } ( P_{\eta } A^{\rm R}_{\eta } ) & = - P_{\eta } Q \Big( \eta \Lambda ^{\rm R} - \Ld A^{\rm NS}_{\eta } , \Lambda ^{\rm R} \Rd _{\ast } - \Ld A^{\rm R} , \Lambda ^{\rm NS} \Rd _{\ast } \Big) , 
\end{align}
where we use a R state $\Lambda ^{\rm R}$, a redefined R gauge parameter, 
\begin{align}
\nonumber 
\Lambda ^{\rm R} \equiv F \Xi \Big( - \lambda ^{\rm R} + \Ld A^{\rm R}_{\eta } , \Lambda ^{\rm NS} \Rd _{\ast } \Big) , 
\end{align}
which belongs to the large Hilbert space $\mathcal{H}$ and has ghost-and-picture number $( -1 | \frac{1}{2} )$. 

In the work of \cite{Kunitomo:2015usa}, all computations of the variation of the action, equations of motion, and gauge invariance heavily depend on the explicit form or properties of the linear map $F$ on $\Psi ^{\rm R}$. 
However, as we will show in the next section, these all computations are derived from WZW-like properties of the Ramond sector: (\ref{R eta}), (\ref{R eta XY}), and (\ref{R d}). 

\section{Wess-Zumino-Witten-like complete action}

We first summarize Wess-Zumino-Witten-like relations of the NS sector and the R sector separately, and show that these relations indeed provide the topological parameter dependence of the action. 
Second, coupling NS and R, we give a Wess-Zumino-Witten-like complete action and prove that the gauge invariance of the action is also derived from the WZW-like relations. 
Lastly, we introduce a notation unifying separately given results of NS and R sectors, and another form of the action, which we call a single functional form.

\subsection{WZW-like structure and $XY$-projection} 

Let $\varphi $ is a dynamical string field. 
We write $\varphi (t)$ for a path satisfying $\varphi (0) = 0$ and $\varphi (1) = \varphi $.

\niu{Neveu-Schwarz sector} 

An NS pure-gauge-like (functional) field ${\cal A}^{\rm NS}_{\eta } = \mathcal{A}^{\rm NS}_{\eta }[ \varphi ]$ is a ghost-and-picture number $(1|-1)$ state satisfying 
\begin{align}
\label{NS pure-gauge-like equation}
\eta {\cal A}^{\rm NS}_{\eta } - \frac{1}{2} \ld {\cal A}^{\rm NS}_{\eta } , {\cal A}^{\rm NS}_{\eta } \rd _{\ast } = 0 . 
\end{align}
Let $d$ be a derivation operator satisfying $\ld d , \eta \rd = 0$, and let $(d_{\rm g} | d_{\rm p})$ be ghost-and-picture number of $d$. 
For example, one can take $d = Q, \partial _{t}, \delta $. 
An NS associated (functional) field ${\cal A}^{\rm NS}_{d} = \mathcal{A}^{\rm NS}_{d} [\varphi ] $ is a ghost-and-picture number $(d_{\rm g} | d_{\rm p} )$ state satisfying 
\begin{align}
\label{NS associated} 
(-)^{d} d {\cal A}^{\rm NS}_{\eta } & = \eta {\cal A}^{\rm NS}_{d} - \ld {\cal A}^{\rm NS}_{\eta } , {\cal A}^{\rm NS}_{d} \rd _{\ast }  
\no 
& \equiv D^{\rm NS}_{\eta } {\cal A}^{\rm NS}_{d} . 
\end{align}
By definition of $(\ref{NS pure-gauge-like equation})$ and $(\ref{NS associated})$, one can check that the relation  
\begin{align} 
\label{NS bianchi} 
D^{\rm NS}_{\eta } \Big( d_{1} {\cal A}^{\rm NS}_{d_{2}} - (-)^{d_{1} d_{2}} d_{2} {\cal A}^{\rm NS}_{d_{1} } - (-)^{d_{1} d_{2}} \Ld {\cal A}^{\rm NS}_{d_{1}} , {\cal A}^{\rm NS}_{d_{2}} \Rd _{\ast } \Big) = 0 
\end{align}
holds when two derivations $d_{1}$ and $d_{2}$ satisfy $\ld d_{1} , d_{2} \rd \equiv d_{1} d _{2} - (-)^{d_{1} d_{2}} d_{2} d_{1} = 0$. 

Utilizing these (functional) fields, an NS action is given by 
\begin{align}
S^{\rm NS} [\varphi ] & = - \int_{0}^{1} dt \, \langle {\cal A}^{\rm NS}_{t} [ \varphi (t)  ] , Q {\cal A}^{\rm NS}_{\eta } [ \varphi (t) ] \rangle  . 
\end{align} 
Note that $A^{\rm NS}_{\eta } [\varphi (t) ]$ is a functional of the path $\varphi (t)$, and $t$-dependence of the action is fixed by that of $\varphi (t)$. 
It is known that the variation of the NS action is given 
\begin{align}
\delta S^{\rm NS} [\varphi ] = - \langle {\cal A}^{\rm NS}_{\delta } [ \varphi ] , \, Q {\cal A}^{\rm NS}_{\eta } [\varphi ] \rangle , 
\end{align}
which we call the topological parameter dependence of WZW-like action. 
See \cite{Berkovits:2004xh, Erler:2015rra, Erler:2015uoa, GM2}. 

\niu{Ramond sector} 

An R pure-gauge-like (functional) field ${\cal A}^{\rm R}_{\eta } = \mathcal{A}^{\rm R}_{\eta } [ \varphi ] $ is a ghost-and-picture number $(1|-\frac{1}{2} )$ state satisfying 
\begin{align}
\label{R pure-gauge-like equation} 
\eta {\cal A}^{\rm R}_{\eta } - \ld {\cal A}^{\rm NS}_{\eta } , {\cal A}^{\rm R}_{\eta } \rd _{\ast } = 0 , 
\end{align}
or equivalently, $D^{\rm NS}_{\eta } {\cal A}^{\rm R}_{\eta } = 0$. 
Let $d$ be a derivation operator satisfying $\ld d , \eta \rd = 0$, and let $(d_{\rm g} | d_{\rm p})$ be ghost-and-picture number of $d$. 
For example, we can take $d = Q , \partial _{t} , \delta$. 
An R associated (functional) field ${\cal A}^{\rm R}_{d} = \mathcal{A}^{\rm R}_{d} [ \varphi ] $ is a ghost-and-picture number $( d_{\rm g} | d_{\rm p} + \frac{1}{2} )$ state satisfying 
\begin{align} 
\label{R associated equation}
(-)^{d} d {\cal A}^{\rm R}_{\eta } & = \eta {\cal A}^{\rm R}_{d} - \ld {\cal A}^{\rm NS}_{\eta } , {\cal A}^{\rm R}_{d} \rd _{\ast } - \ld {\cal A}^{\rm R}_{\eta } , {\cal A}^{\rm NS}_{d} \rd _{\ast } ,
\end{align}
namely, $(-)^{d} d {\cal A}^{\rm R}_{\eta } = D^{\rm NS}_{\eta } {\cal A}^{\rm R}_{d} - \ld {\cal A}^{\rm R}_{\eta } , {\cal A}^{\rm NS}_{d} \rd _{\ast }$. 

\vspace{2mm} 

As we will show, utilizing these fields and assuming $XY$-projection invariance of $P_{\eta } {\cal A}^{\rm R}_{\eta }$ 
\begin{align}
\label{XY=1}
X Y ( P_{\eta } {\cal A}^{\rm R}_{\eta } ) = P_{\eta } {\cal A}^{\rm R}_{\eta } , 
\end{align}
or equivalently $XY (\xi \mathcal{A}^{\rm R}_{\eta } ) = \xi \mathcal{A}^{\rm R}_{\eta }$ when $P_{\eta } = \eta \xi $ holds, one can construct a gauge invariant action Wess-Zumino-Witten-likely, whose parameter dependence is topological. 
We propose that an R action is given by 
\begin{align}
\nonumber 
S^{\rm R} [\varphi ] & = - \int_{0}^{1} dt \Big( \langle \xi Y \partial _{t} ( P_{\eta } {\cal A}^{\rm R}_{\eta } [\varphi (t) ]) , Q {\cal A}^{\rm R}_{\eta } [\varphi (t) ]\rangle + \langle {\cal A}^{\rm NS}_{t} [\varphi (t) ] , m_{2} ( {\cal A}^{\rm R}_{\eta } [\varphi (t) ] , {\cal A}^{\rm R}_{\eta } [\varphi (t)] ) \rangle \Big) . 
\end{align} 
This $S^{\rm R}$ is Wess-Zumino-Witten-like. 
In other words, $S^{\rm R}$ has topological $t$-dependence 
\begin{align} 
\label{delta Swzw}
\delta S^{\rm R} [\varphi ] =  - \langle \xi Y \delta ( P_{\eta } {\cal A}^{\rm R}_{\eta } [\varphi ] ) , Q {\cal A}^{\rm R}_{\eta } [\varphi ] \rangle - \langle {\cal A}^{\rm NS}_{\delta } [\varphi ] , m_{2}( {\cal A}^{\rm R}_{\eta } [\varphi ] , {\cal A}^{\rm R}_{\eta } [\varphi ] ) \rangle . 
\end{align} 

\niu{Topological $t$-dependence of $S^{\rm R}$}

First, we consider the variation of the first term of $S^{\rm R}$. 
This term consists of two ingredients: 
\begin{align} 
\nonumber 
\int_{0}^{1} dt \, \langle \xi Y \partial _{t} ( P_{\eta } {\cal A}^{\rm R}_{\eta } ) , Q {\cal A}^{\rm R}_{\eta } \rangle  = \int_{0}^{1} dt \, \Big( \langle \xi Y \partial _{t} ( P_{\eta } {\cal A}^{\rm R}_{\eta } ) , Q ( P_{\eta } {\cal A}^{\rm R}_{\eta } ) \rangle - \langle \partial _{t} ( P_{\eta } {\cal A}^{\rm R}_{\eta } ) , {\cal A}^{\rm R}_{\eta } \rangle \Big) . 
\end{align} 
We can quickly find that {\it the first part has topological $t$-dependence} 
\begin{align}
\delta \langle \xi Y \partial _{t} ( P_{\eta } {\cal A}^{\rm R}_{\eta } ) , Q ( P_{\eta } {\cal A}^{\rm R}_{\eta } ) \rangle & = \langle \frac{\partial }{\partial t} \big{\{ } \xi Y \delta ( P_{\eta } {\cal A}^{\rm R}_{\eta } ) \big{\} } , Q ( P_{\eta } {\cal A}^{\rm R}_{\eta } ) \rangle + \langle  \xi Y \partial _{t} ( P_{\eta } {\cal A}^{\rm R}_{\eta } ) , Q \delta ( P_{\eta } {\cal A}^{\rm R}_{\eta } ) \rangle 
\no \nonumber 
& = \frac{\partial }{\partial t} \langle \xi Y \delta ( P_{\eta } {\cal A}^{\rm R}_{\eta } ) , Q ( P_{\eta } {\cal A}^{\rm R}_{\eta } ) \rangle   
\end{align}
since using (\ref{XY=1}), $\xi Q - X = - Q \xi $, and $\langle \partial _{t}( P_{\eta } {\cal A}^{\rm R}_{\eta }) , \delta ( P_{\eta } {\cal A}^{\rm R}_{\eta }) \rangle = 0$, the following relation holds 
\begin{align}
\langle  \xi Y \partial _{t} ( P_{\eta } {\cal A}^{\rm R}_{\eta } ) , Q \delta ( P_{\eta } {\cal A}^{\rm R}_{\eta } ) \rangle & = \langle ( \xi Q - X) Y \partial _{t} ( P_{\eta } {\cal A}^{\rm R}_{\eta } ) ,  X Y \delta ( P_{\eta } {\cal A}^{\rm R}_{\eta } ) \rangle 
\no 
& = \langle Q \partial _{t} ( P_{\eta } {\cal A}^{\rm R}_{\eta } ) , \xi Y \delta ( P_{\eta } {\cal A}^{\rm R}_{\eta } ) \rangle 
\no \nonumber 
& = \langle  \xi Y \delta ( P_{\eta } {\cal A}^{\rm R}_{\eta } ) , \frac{\partial }{\partial t} \big{\{ } Q ( P_{\eta } {\cal A}^{\rm R}_{\eta } ) \big{\} } \rangle. 
\end{align} 
Note however that {\it the variation of the second ingredient provides an extra term} 
\begin{align} 
\delta \langle \partial _{t} ( P_{\eta } {\cal A}^{\rm R}_{\eta } ) , {\cal A}^{\rm R}_{\eta } \rangle & =  \langle \frac{\partial }{\partial t} \big{\{ } \delta ( P_{\eta } {\cal A}^{\rm R}_{\eta } ) \big{\} } , {\cal A}^{\rm R}_{\eta } \rangle +  \langle \partial _{t} ( P_{\eta } {\cal A}^{\rm R}_{\eta } ) , \delta {\cal A}^{\rm R}_{\eta } \rangle 
\no 
& = \frac{\partial }{\partial t} \langle \delta ( P_{\eta } {\cal A}^{\rm R}_{\eta } ) , {\cal A}^{\rm R}_{\eta } \rangle -  \langle \delta ( P_{\eta } {\cal A}^{\rm R}_{\eta } ) , \partial _{t} {\cal A}^{\rm R}_{\eta } \rangle +  \langle \partial _{t} ( P_{\eta } {\cal A}^{\rm R}_{\eta } ) , \delta {\cal A}^{\rm R}_{\eta } \rangle 
\no \nonumber 
& = \frac{\partial }{\partial t} \langle \delta ( P_{\eta } {\cal A}^{\rm R}_{\eta } ) , {\cal A}^{\rm R}_{\eta } \rangle -  \langle \delta  {\cal A}^{\rm R}_{\eta } , \partial _{t} {\cal A}^{\rm R}_{\eta } \rangle .  
\end{align}
Here, we used $P_{\eta } + P_{\xi } =1$ and $\langle \partial _{t} ( P_{\eta } {\cal A}^{\rm R}_{\eta } ) , \delta {\cal A}^{\rm R}_{\eta } \rangle = \langle \partial _{t} {\cal A}^{\rm R}_{\eta } , \delta ( P_{\xi } {\cal A}^{\rm R}_{\eta } ) \rangle = - \langle \delta ( P_{\xi } {\cal A}^{\rm R}_{\eta } ) , \partial _{t} {\cal A}^{\rm R}_{\eta } \rangle$. 
As a result, the variation of the first term of $S^{\rm R}$ is given by 
\begin{align} 
\label{1st term}
\delta \int_{0}^{1} dt \, \langle \xi Y \partial _{t} ( P_{\eta } {\cal A}^{\rm R}_{\eta } ) , Q {\cal A}^{\rm R}_{\eta } \rangle = \langle \xi Y \delta ( P_{\eta } {\cal A}^{\rm R}_{\eta } ) , Q {\cal A}^{\rm R}_{\eta } \rangle + \int_{0}^{1} dt \, \langle \delta {\cal A}^{\rm R}_{\eta } , \partial _{t} {\cal A}^{\rm R}_{\eta } \rangle . 
\end{align} 

Second, we compute the variation of the second term of $S^{\rm R}$. 
Using (\ref{NS bianchi}), (\ref{R associated equation}) for $d = \partial _{t} , \delta $, and Jacobi identities of the commutator, we obtain 
\begin{align} 
\label{2nd term}
\delta \langle {\cal A}^{\rm NS}_{t} , m_{2} ( {\cal A}^{\rm R}_{\eta } , {\cal A}^{\rm R}_{\eta } ) \rangle & =  \langle \delta {\cal A}^{\rm NS}_{t} , m_{2} ( {\cal A}^{\rm R}_{\eta } , {\cal A}^{\rm R}_{\eta } ) \rangle +  \langle {\cal A}^{\rm NS}_{t} , \ld {\cal A}^{\rm R}_{\eta } , \delta {\cal A}^{\rm R}_{\eta } \rd _{\ast } \rangle
\no 
& =  \langle \partial {\cal A}^{\rm NS}_{\delta } , m_{2} ( {\cal A}^{\rm R}_{\eta } , {\cal A}^{\rm R}_{\eta } ) \rangle + \langle \ld {\cal A}^{\rm NS}_{t} \! \! , {\cal A}^{\rm NS}_{\delta } \rd _{\ast } , m_{2} ( {\cal A}^{\rm R}_{\eta } , {\cal A}^{\rm R}_{\eta } ) \rangle +  \langle {\cal A}^{\rm NS}_{t} , \ld {\cal A}^{\rm R}_{\eta } , \delta {\cal A}^{\rm R}_{\eta } \rd _{\ast } \rangle 
\no 
& = \frac{\partial }{\partial t}  \langle {\cal A}^{\rm NS}_{\delta } , m_{2} ( {\cal A}^{\rm R}_{\eta } , {\cal A}^{\rm R}_{\eta } ) \rangle
- \langle {\cal A}^{\rm NS}_{\delta } , \ld {\cal A}^{\rm R}_{\eta } , \partial _{t} {\cal A}^{\rm R}_{\eta } \rd _{\ast} \rangle 
\no 
& \hspace{20mm} - \langle \ld {\cal A}^{\rm NS}_{\delta } , {\cal A}^{\rm NS}_{t} \rd _{\ast } , m_{2} ( {\cal A}^{\rm R}_{\eta } , {\cal A}^{\rm R}_{\eta } ) \rangle +  \langle  \delta {\cal A}^{\rm R}_{\eta }  ,  \ld {\cal A}^{\rm R}_{\eta } , {\cal A}^{\rm NS}_{t} \rd _{\ast } \rangle 
\no 
& = \frac{\partial }{\partial t}  \langle {\cal A}^{\rm NS}_{\delta } , m_{2} ( {\cal A}^{\rm R}_{\eta } , {\cal A}^{\rm R}_{\eta } ) \rangle
- \frac{1}{2} \langle \ld {\cal A}^{\rm NS}_{\delta } , {\cal A}^{\rm NS}_{t} \rd _{\ast } , \ld  {\cal A}^{\rm R}_{\eta } , {\cal A}^{\rm R}_{\eta } \rd _{\ast }\rangle
\no \nonumber 
&  \hspace{30mm} + \langle \ld {\cal A}^{\rm R}_{\eta } , {\cal A}^{\rm NS}_{\delta } \rd _{\ast } , \partial _{t} {\cal A}^{\rm R}_{\eta } \rangle 
+  \langle  \delta {\cal A}^{\rm R}_{\eta }  ,  \ld {\cal A}^{\rm R}_{\eta } , {\cal A}^{\rm NS}_{t} \rd _{\ast } \rangle 
\no 
& = \frac{\partial }{\partial t}  \langle {\cal A}^{\rm NS}_{\delta } , m_{2} ( {\cal A}^{\rm R}_{\eta } , {\cal A}^{\rm R}_{\eta } ) \rangle 
- \langle \ld {\cal A}^{\rm R}_{\eta } , {\cal A}^{\rm NS}_{\delta } \rd _{\ast } ,  \ld {\cal A}^{\rm R}_{\eta } , {\cal A}^{\rm NS}_{t} \rd _{\ast } \rangle 
\no \nonumber 
& \hspace{25mm} + \Big( \langle D^{\rm NS}_{\eta } {\cal A}^{\rm R}_{\delta }  ,   \ld {\cal A}^{\rm R}_{\eta } , {\cal A}^{\rm NS}_{t} \rd _{\ast } \rangle + \langle \ld {\cal A}^{\rm R}_{\eta } , {\cal A}^{\rm NS}_{\delta } \rd _{\ast } ,  D^{\rm NS}_{\eta } {\cal A}^{\rm R}_{t}  \rd _{\ast } \rangle \Big) 
\end{align}
In particular, from the forth line to the last line, we applied  
\begin{align}
\langle \ld {\cal A}^{\rm R}_{\eta } , {\cal A}^{\rm NS}_{\delta } \rd _{\ast } , \partial _{t} {\cal A}^{\rm R}_{\eta } \rangle  & = 
\langle \ld {\cal A}^{\rm R}_{\eta } , {\cal A}^{\rm NS}_{\delta } \rd _{\ast } , D^{\rm NS}_{\eta } {\cal A}^{\rm R}_{t} - \ld {\cal A}^{\rm R}_{\eta } , {\cal A}^{\rm NS}_{t} \rd _{\ast } \rangle 
\no 
& = \langle \ld {\cal A}^{\rm R}_{\eta } , {\cal A}^{\rm NS}_{\delta } \rd _{\ast } ,  D^{\rm NS}_{\eta } {\cal A}^{\rm R}_{t}  \rd _{\ast } \rangle 
- \langle \ld {\cal A}^{\rm R}_{\eta } , {\cal A}^{\rm NS}_{\delta } \rd _{\ast } ,  \ld {\cal A}^{\rm R}_{\eta } , {\cal A}^{\rm NS}_{t} \rd _{\ast } \rangle . 
\no[2mm]  
\langle  \delta {\cal A}^{\rm R}_{\eta } , \ld {\cal A}^{\rm R}_{\eta } , {\cal A}^{\rm NS}_{t} \rd _{\ast } \rangle & = \langle D^{\rm NS}_{\eta } {\cal A}^{\rm R}_{\delta } - \ld {\cal A}^{\rm R}_{\eta } , {\cal A}^{\rm NS}_{\delta } \rd _{\ast } , \ld {\cal A}^{\rm R}_{\eta } , {\cal A}^{\rm NS}_{t} \rd _{\ast } \rangle  
\no  
& = \langle \ld {\cal A}^{\rm R}_{\eta } , {\cal A}^{\rm NS}_{\delta } \rd _{\ast } ,  D^{\rm NS}_{\eta } {\cal A}^{\rm R}_{t} \rangle 
-  \langle \ld {\cal A}^{\rm R}_{\eta } , {\cal A}^{\rm NS}_{\delta } \rd _{\ast } ,  \ld {\cal A}^{\rm R}_{\eta } , {\cal A}^{\rm NS}_{t} \rd _{\ast } \rangle . 
\no[2mm] \nonumber 
- \frac{1}{2} \langle \ld {\cal A}^{\rm NS}_{\delta } , {\cal A}^{\rm NS}_{t} \rd _{\ast } , \ld  {\cal A}^{\rm R}_{\eta } , {\cal A}^{\rm R}_{\eta } \rd _{\ast } \rangle & = \frac{1}{2} \langle {\cal A}^{\rm NS}_{\delta } , \Ld  \ld  {\cal A}^{\rm R}_{\eta } , {\cal A}^{\rm R}_{\eta } \rd _{\ast } , {\cal A}^{\rm NS}_{t}  \Rd _{\ast } \rangle
= - \langle {\cal A}^{\rm NS}_{\delta } , \Ld  \ld  {\cal A}^{\rm R}_{\eta } , {\cal A}^{\rm R}_{t} \rd _{\ast } , {\cal A}^{\rm NS}_{\eta }  \Rd _{\ast } \rangle
\no \nonumber 
& = \langle \ld {\cal A}^{\rm R}_{\eta } , {\cal A}^{\rm NS}_{\delta } \rd _{\ast } ,  \ld {\cal A}^{\rm R}_{\eta } , {\cal A}^{\rm NS}_{t} \rd _{\ast } \rangle . 
\end{align} 
As a result, the variation of the second term of $S^{\rm R}$ is given by 
\begin{align}
\delta \int_{0}^{1} dt \, \langle {\cal A}^{\rm NS}_{t} , m_{2} ( {\cal A}^{\rm R}_{\eta } , {\cal A}^{\rm NS}_{\eta } ) \rangle = \langle {\cal A}^{\rm NS}_{\delta } , m_{2} ( {\cal A}^{\rm R}_{\eta } , {\cal A}^{\rm R}_{\eta } ) \rangle
- \int_{0}^{1} dt \, \langle \delta {\cal A}^{\rm R}_{\eta } ,  \partial _{t} {\cal A}^{\rm R}_{\eta } \rangle  
\end{align} 
because of the following relation 
\begin{align}
\langle \delta {\cal A}^{\rm R}_{\eta } , \partial _{t} {\cal A}^{\rm R}_{\eta } \rangle & = \langle D^{\rm NS}_{\eta } {\cal A}^{\rm R}_{\delta } - \ld {\cal A}^{\rm R}_{\eta } , {\cal A}^{\rm NS}_{\delta } \rd _{\ast } ,  D^{\rm NS}_{\eta } {\cal A}^{\rm R}_{t} - \ld {\cal A}^{\rm R}_{\eta } , {\cal A}^{\rm NS}_{t} \rd \rangle 
\no \nonumber 
& = \langle \ld {\cal A}^{\rm R}_{\eta } , {\cal A}^{\rm NS}_{\delta } \rd _{\ast } ,  \ld {\cal A}^{\rm R}_{\eta } , {\cal A}^{\rm NS}_{t} \rd _{\ast } \rangle - \Big( \langle D^{\rm NS}_{\eta } {\cal A}^{\rm R}_{\delta }  ,   \ld {\cal A}^{\rm R}_{\eta } , {\cal A}^{\rm NS}_{t} \rd _{\ast } \rangle + \langle \ld {\cal A}^{\rm R}_{\eta } , {\cal A}^{\rm NS}_{\delta } \rd _{\ast } ,  D^{\rm NS}_{\eta } {\cal A}^{\rm R}_{t} \rangle \Big) . 
\end{align}
Hence, (\ref{1st term}) plus (\ref{2nd term}) provides that $S^{\rm R}$ has topological $t$-dependence (\ref{delta Swzw}).

\subsection{WZW-like complete action} 

We propose a Wess-Zumino-Witten complete action and show its gauge invariance on the basis of WZW-like relations (\ref{NS pure-gauge-like equation} - \ref{NS bianchi}) and (\ref{R pure-gauge-like equation} - \ref{XY=1}).

\niu{Action and equations of motion}

We propose that a Wess-Zumino-Witten-like complete action is given by 
\begin{align}
S_{\rm wzw} [\varphi ] & \equiv S^{\rm NS} [\varphi ]+ S^{\rm R} [\varphi ]  
\no 
& =  - \int_{0}^{1} dt \Big( \langle \xi Y \partial _{t} ( P_{\eta } {\cal A}^{\rm R}_{\eta } ) , Q {\cal A}^{\rm R}_{\eta } \rangle + \langle {\cal A}^{\rm NS}_{t} , Q {\cal A}^{\rm NS}_{\eta } + m_{2} ( {\cal A}^{\rm R}_{\eta } , {\cal A}^{\rm R}_{\eta } ) \rangle \Big) . 
\end{align} 
Here, $\varphi $ is a dynamical string field of the theory and $\mathcal{A}^{\rm NS/R}_{\eta } = \mathcal{A}^{\rm NS/R}_{\eta }[\varphi (t) ]$ and $\mathcal{A}^{\rm NS}_{t} = \mathcal{A}^{\rm NS}_{t}[\varphi (t) ]$ are functionals of the path $\varphi (t)$ satisfying $\varphi (0) = 0$ and $\varphi (1) = \varphi $. 
Since $S^{\rm NS}$ and $S^{\rm R}$ have topological $t$-dependence, the variation of the action $S_{\rm WZW}$ is given by 
\begin{align}
- \delta S_{\rm wzw} [\varphi ] = \langle \xi Y \delta ( P_{\eta } {\cal A}^{\rm R}_{\eta } ) , Q {\cal A}^{\rm R}_{\eta } \rangle + \langle {\cal A}^{\rm NS}_{\delta } , Q {\cal A}^{\rm NS}_{\eta } + m_{2} ( {\cal A}^{\rm R}_{\eta } , {\cal A}^{\rm R}_{\eta } ) \rangle , 
\end{align} 
where $\mathcal{A}^{\rm NS/R}_{\eta } = \mathcal{A}^{\rm NS/R}_{\eta }[\varphi ]$ and $\mathcal{A}^{\rm NS}_{t} = \mathcal{A}^{\rm NS}_{\delta }[\varphi ]$ are functionals of the dynamical string field $\varphi $, which is the end point of the path $\varphi (1)$. 
We therefore obtain the equations of motion 
\begin{align}
\label{NS EOM}
{\rm NS} \, : & \hspace{5mm} Q {\cal A}^{\rm NS} [\varphi ] + m_{2} ( {\cal A}^{\rm R}_{\eta } [\varphi ] , {\cal A}^{\rm R}_{\eta } [\varphi ] ) = 0 , 
\\ \label{R EOM}
{\rm R} \, : & \hspace{5mm} P_{\eta } \big( Q {\cal A}^{\rm R}_{\eta } [\varphi ] \big) = 0 . 
\end{align} 

Let $\Lambda ^{\rm NS}$, $\Lambda ^{\rm R}$, and $\Omega ^{\rm NS}$ be NS, R, and NS gauge parameter fields which have ghost-and-picture number $(-1|0)$, $(-1|\frac{1}{2})$, and $(-1|1)$, respectively. 
These $\Lambda ^{\rm NS}$, $\Lambda ^{\rm R}$, and $\Omega ^{\rm NS}$ all belong to the large Hilbert space. 
The action is invariant under two types of gauge transformations: 
the gauge transformations parametrized by $\Lambda = \Lambda ^{\rm NS} + \Lambda ^{\rm R}$  
\begin{align} 
\label{delta Lambda NS}
{\cal A}^{\rm NS}_{\delta _{\Lambda }} & = Q \Lambda ^{\rm NS} + \Ld {\cal A}^{\rm R}_{\eta } , \Lambda ^{\rm R} \Rd _{\ast }, 
\\ 
\label{delta Lambda R}
\delta _{\Lambda } ( P_{\eta } {\cal A}^{\rm R}_{\eta } ) & = - P_{\eta } Q \Big( \eta \Lambda ^{\rm R} - \Ld {\cal A}^{\rm NS}_{\eta } , \Lambda ^{\rm R} \Rd _{\ast } - \Ld {\cal A}^{\rm R}_{\eta } , \Lambda ^{\rm NS} \Rd _{\ast } \Big) , 
\end{align}
and the gauge transformations parametrized by $\Omega = \Omega ^{\rm NS}$ 
\begin{align}
\label{delta Omega}
{\cal A}^{\rm NS}_{\delta _{\Omega } }  = \eta \Omega ^{\rm NS} - \Ld {\cal A}^{\rm NS}_{\eta } , \Omega ^{\rm NS} \Rd _{\ast } , 
\hspace{5mm}  
\delta _{\Omega } ( P_{\eta } {\cal A}^{\rm R}_{\eta } ) = 0 . 
\end{align}

\niu{$\Lambda$-gauge invariance}

The $\Lambda $-gauge transformations of the action is given by 
\begin{align}
- \label{delta Lambda}
\delta _{\Lambda } S_{\rm wzw} = \langle \xi Y \delta _{\Lambda } ( P_{\eta } {\cal A}^{\rm R}_{\eta } ) , Q {\cal A}^{\rm R}_{\eta } \rangle + \langle {\cal A}^{\rm NS}_{\delta _{\Lambda }} , Q {\cal A}^{\rm NS}_{\eta } + m_{2} ( {\cal A}^{\rm R}_{\eta } , {\cal A}^{\rm R}_{\eta } ) \rangle . 
\end{align} 
We show that $\delta _{\Lambda } S_{\rm WZW } = 0$ with $\Lambda$-gauge transformations of fields 
\begin{align} 
{\cal A}^{\rm NS}_{\delta _{\Lambda }} & = Q \Lambda ^{\rm NS} + \Ld {\cal A}^{\rm R}_{\eta } , \Lambda ^{\rm R} \Rd _{\ast } , 
\no \nonumber 
\delta _{\Lambda } ( P_{\eta } {\cal A}^{\rm R}_{\eta } ) & = - P_{\eta } Q \Big( \eta \Lambda ^{\rm R} - \Ld {\cal A}^{\rm NS}_{\eta } , \Lambda ^{\rm R} \Rd _{\ast } - \Ld {\cal A}^{\rm R}_{\eta } , \Lambda ^{\rm NS} \Rd _{\ast } \Big) , 
\end{align} 
where $\Lambda ^{\rm NS}$ is an NS gauge parameter carrying ghost-and-picture number $(-1 | 0)$ and $\Lambda ^{\rm R}$ is a R gauge parameter carrying ghost-and-picture number $(-1 | \frac{1}{2})$.  
Note that these $\Lambda ^{\rm NS}$ and $\Lambda ^{\rm R}$ belong to the large Hilbert space $\mathcal{H}$ but $\delta _{\Lambda } (P_{\eta } {\cal A}^{\rm R}_{\eta } )$ has to be in the restricted one $\mathcal{H}_{R}$. 

First, we consider the first term of (\ref{delta Lambda}) with (\ref{delta Lambda R}). 
This term consists of two ingredients, 
\begin{align}
\label{R part of delta Lambda}
\langle \xi Y \delta _{\Lambda } ( P_{\eta } {\cal A}^{\rm R}_{\eta } ) , Q {\cal A}^{\rm R}_{\eta } \rangle & = \langle \xi Y \delta _{\Lambda } ( P_{\eta } {\cal A}^{\rm R}_{\eta } ) , \, Q (P_{\eta } + P_{\xi } ) {\cal A}^{\rm R}_{\eta } \rangle 
\no 
& = \langle \xi Y \delta _{\Lambda } ( P_{\eta } {\cal A}^{\rm R}_{\eta } ) , Q X Y ( P_{\eta } {\cal A}^{\rm R}_{\eta } ) \rangle - \langle \delta _{\Lambda } ( P_{\eta } {\cal A}^{\rm R}_{\eta } ) ,  P_{\xi } {\cal A}^{\rm R}_{\eta } \rangle 
\no 
& = \langle \xi Q \delta _{\Lambda } ( P_{\eta } {\cal A}^{\rm R}_{\eta } ) , Y ( P_{\eta } {\cal A}^{\rm R}_{\eta } ) \rangle - \langle \delta _{\Lambda } ( P_{\eta } {\cal A}^{\rm R}_{\eta } ) ,  P_{\xi } {\cal A}^{\rm R}_{\eta } \rangle . 
\end{align}
Here, we used $\delta ( P_{\eta } {\cal A}^{\rm R}_{\eta } ) = P_{\eta } ( \delta P_{\eta } {\cal A}^{\rm R}_{\eta } ) $, $X Y ( \delta P_{\eta } {\cal A}^{\rm R}_{\eta } ) = \delta ( X Y P_{\eta } {\cal A}^{\rm R}_{\eta })$, and $X Y ( P_{\eta } {\cal A}^{\rm R}_{\eta } ) = P_{\eta } {\cal A}^{\rm R}_{\eta }$. 
Since the first ingredient of (\ref{R part of delta Lambda}) with $\Lambda $-gauge transformations (\ref{delta Lambda R}) becomes 
\begin{align}
\langle \xi Q \delta _{\Lambda } ( P_{\eta } {\cal A}^{\rm R}_{\eta } ) , Y ( P_{\eta } {\cal A}^{\rm R}_{\eta } ) \rangle & 
= - \langle \xi Q P_{\eta } Q \big( D^{\rm NS}_{\eta } \Lambda ^{\rm R} - \Ld {\cal A}^{\rm R}_{\eta } , \Lambda ^{\rm NS} \Rd _{\ast } \big) , Y ( P_{\eta } {\cal A}^{\rm R}_{\eta } ) \rangle 
\no 
& = \langle P_{\xi } Q \xi Q \big( D^{\rm NS}_{\eta } \Lambda ^{\rm R} - \Ld {\cal A}^{\rm R}_{\eta } , \Lambda ^{\rm NS} \Rd _{\ast } \big) , Y ( P_{\eta } {\cal A}^{\rm R}_{\eta } ) \rangle 
\no \nonumber 
& = \langle Q \big( D^{\rm NS}_{\eta } \Lambda ^{\rm R} - \Ld {\cal A}^{\rm R}_{\eta } , \Lambda ^{\rm NS} \Rd _{\ast } \big) , P_{\eta } {\cal A}^{\rm R}_{\eta } \rangle 
\end{align}
and the second ingredient of (\ref{R part of delta Lambda}) with $\Lambda $-gauge transformations (\ref{delta Lambda R}) becomes 
\begin{align}
- \langle \delta _{\Lambda } ( P_{\eta } {\cal A}^{\rm R}_{\eta } ) ,  P_{\xi } {\cal A}^{\rm R}_{\eta } \rangle & =  \langle P_{\eta } Q \big( D^{\rm NS}_{\eta } \Lambda ^{\rm R} - \Ld {\cal A}^{\rm R}_{\eta } , \Lambda ^{\rm NS} \Rd _{\ast } \big) , P_{\xi } {\cal A}^{\rm R}_{\eta } \rangle 
\no \nonumber 
& =  \langle Q \big( D^{\rm NS}_{\eta } \Lambda ^{\rm R} - \Ld {\cal A}^{\rm R}_{\eta } , \Lambda ^{\rm NS} \Rd _{\ast } \big) , P_{\xi } {\cal A}^{\rm R}_{\eta } \rangle , 
\end{align}
we obtain 
\begin{align}
\label{R delta Lambda 1}
\langle \xi Y \delta _{\Lambda } ( P_{\eta } {\cal A}^{\rm R}_{\eta } ) , Q {\cal A}^{\rm R}_{\eta } \rangle & = \langle \xi Q \delta _{\Lambda } ( P_{\eta } {\cal A}^{\rm R}_{\eta } ) , Y ( P_{\eta } {\cal A}^{\rm R}_{\eta } ) \rangle - \langle \delta _{\Lambda } ( P_{\eta } {\cal A}^{\rm R}_{\eta } ) ,  P_{\xi } {\cal A}^{\rm R}_{\eta } \rangle 
\no 
& = \langle Q \big( D^{\rm NS}_{\eta } \Lambda ^{\rm R} - \Ld {\cal A}^{\rm R}_{\eta } , \Lambda ^{\rm NS} \Rd _{\ast } \big) ,  P_{\eta } {\cal A}^{\rm R}_{\eta } + P_{\xi } {\cal A}^{\rm R}_{\eta } \rangle 
\no 
&= - \langle D^{\rm NS}_{\eta } \Lambda ^{\rm R} - \Ld {\cal A}^{\rm R}_{\eta } , \Lambda ^{\rm NS} \Rd _{\ast } , Q {\cal A}^{\rm R}_{\eta } \rangle . 
\end{align}

Next, we compute the second term of (\ref{delta Lambda}) with (\ref{delta Lambda NS}). 
Using $Q^{2} = 0$, $\Ld \ld {\cal A}^{\rm R}_{\eta } , {\cal A}^{\rm R}_{\eta } \rd _{\ast } , {\cal A}^{\rm R}_{\eta } \Rd _{\ast }  = 0$, and $D^{\rm NS}_{\eta } {\cal A}^{\rm R}_{\eta } = 0$, we quickly find that 
\begin{align}
\langle {\cal A}^{\rm NS}_{\delta _{\Lambda }} , Q {\cal A}^{\rm NS}_{\eta } + m_{2} ( {\cal A}^{\rm R}_{\eta } , {\cal A}^{\rm R}_{\eta } ) \rangle & = \langle Q \Lambda ^{\rm NS} + \Ld {\cal A}^{\rm R}_{\eta } , \Lambda ^{\rm R} \Rd _{\ast } , Q {\cal A}^{\rm NS}_{\eta } + m_{2} ( {\cal A}^{\rm R}_{\eta } , {\cal A}^{\rm R}_{\eta } ) \rangle
\no 
& = \langle \Ld {\cal A}^{\rm R}_{\eta } , \Lambda ^{\rm R} \Rd _{\ast } , Q {\cal A}^{\rm NS}_{\eta } \rangle +  \langle Q \Lambda ^{\rm NS} , m_{2} ( {\cal A}^{\rm R}_{\eta } , {\cal A}^{\rm R}_{\eta } ) \rangle 
\no 
& = - \langle  \Ld {\cal A}^{\rm R}_{\eta } , \Lambda ^{\rm R} \Rd , D^{\rm NS}_{\eta } {\cal A}^{\rm NS}_{Q}  \rangle - \langle \Lambda ^{\rm NS} , \Ld {\cal A}^{\rm R}_{\eta } , Q {\cal A}^{\rm R}_{\eta } \Rd _{\ast }  \rangle
\no \nonumber 
& = \langle  D^{\rm NS}_{\eta } \Lambda ^{\rm R} ,  \Ld {\cal A}^{\rm R}_{\eta } , {\cal A}^{\rm NS}_{Q} \Rd _{\ast } \rangle - \langle \Ld {\cal A}^{\rm R}_{\eta } , \Lambda ^{\rm NS}  \Rd _{\ast } , Q {\cal A}^{\rm R}_{\eta }  \rangle . 
\end{align} 
The property (\ref{R associated equation}) of the R pure-gauge-like field $- Q {\cal A}^{\rm R}_{\eta } = D^{\rm NS}_{\eta } {\cal A}^{\rm R}_{Q} - \Ld {\cal A}^{\rm R}_{\eta } , {\cal A}^{\rm NS}_{Q} \Rd _{\ast } $ gives 
\begin{align}  
\nonumber 
D^{NS}_{\eta } \Big( Q {\cal A}^{\rm R}_{\eta } - \Ld {\cal A}^{\rm R}_{\eta } , {\cal A}^{\rm NS}_{Q} \Rd _{\ast } \Big) = 0 . 
\end{align}
Hence, we obtain 
\begin{align} 
\label{R delta Lambda 2} 
\langle {\cal A}^{\rm NS}_{\delta _{\Lambda }} , Q {\cal A}^{\rm NS}_{\eta } + m_{2} ( {\cal A}^{\rm R}_{\eta } , {\cal A}^{\rm R}_{\eta } ) \rangle = \langle D^{\rm NS}_{\eta } \Lambda ^{\rm R} - \Ld {\cal A}^{\rm R}_{\eta } , \Lambda ^{\rm NS} \Rd _{\ast } , Q {\cal A}^{\rm R}_{\eta } \rangle , 
\end{align}
which just cancels (\ref{R delta Lambda 1}), and we conclude $ \delta _{\Lambda} S_{\rm wzw} = (\ref{R delta Lambda 1}) + (\ref{R delta Lambda 2}) = 0$ with (\ref{delta Lambda NS}) and (\ref{delta Lambda R}).

\niu{$\Omega $-gauge invariance} 

The $\Omega $-gauge transformation of the action $S_{\rm wzw}$ is given by 
\begin{align}
- \delta _{\Omega } S_{\rm wzw} = \langle \xi Y \delta _{\Omega } ( P_{\eta } {\cal A}^{\rm R}_{\eta } ) , Q {\cal A}^{\rm R}_{\eta } \rangle + \langle {\cal A}^{\rm NS}_{\delta _{\Omega }} , Q {\cal A}^{\rm NS}_{\eta } + m_{2} ( {\cal A}^{\rm R}_{\eta } , {\cal A}^{\rm R}_{\eta } ) \rangle . 
\end{align} 
One can show that $\delta _{\Omega } S_{\rm wzw} = 0$ with $\Omega $-gauge transformations of fields 
\begin{align}
\label{delta Omega NS} 
{\cal A}^{\rm NS}_{\delta _{\Omega } }  & = \eta \Omega ^{\rm NS} - \Ld {\cal A}^{\rm NS}_{\eta } , \Omega ^{\rm NS} \Rd _{\ast } - \Ld {\cal A}^{\rm R}_{\eta } , \Omega ^{\rm R} \Rd _{\ast } , 
\\ 
\label{delta Omega R}
\delta _{\Omega } ( P_{\eta } {\cal A}^{\rm R}_{\eta } ) & = - P_{\eta } Q \Big( \eta \Omega ^{\rm R} - \Ld {\cal A}^{\rm NS}_{\eta } , \Omega ^{\rm R} \Rd _{\ast } \Big) ,  
\end{align}
where $\Omega ^{\rm NS}$ is an NS gauge parameter carrying ghost-and-picture number $(-1 | 1)$, $\Omega ^{\rm R}$ is a R gauge parameter carrying ghost-and-picture number $(-1 | \frac{1}{2})$, and both $\Omega ^{\rm NS}$ and $\Omega ^{\rm R}$ belong to the large Hilbert space. 
Note, however, that since R gauge parameters $\Omega ^{\rm R}$ and $\Lambda ^{\rm R}$ have the same ghost-and-picture number $(-1 | \frac{1}{2} )$, we can not distinguish these two parameters. 
As a result, $\Omega ^{\rm R}$-gauge transformation is absorbed into $\Lambda ^{\rm R}$-gauge transformation (\ref{delta Lambda R}) and $\Omega $-gauge transformations (\ref{delta Omega NS}) and (\ref{delta Omega R}) reduces to (\ref{delta Omega}): 
\begin{align}
\nonumber 
{\cal A}^{\rm NS}_{\delta _{\Omega } }  = \eta \Omega ^{\rm NS} - \Ld {\cal A}^{\rm NS}_{\eta } , \Omega ^{\rm NS} \Rd _{\ast } , 
\hspace{5mm}  
\delta _{\Omega } ( P_{\eta } {\cal A}^{\rm R}_{\eta } ) = 0 . 
\end{align} 
Then, using $Q {\cal A}^{\rm NS}_{\eta } = - D^{\rm NS}_{\eta } {\cal A}^{\rm NS}_{Q}$ and $D^{\rm NS}_{\eta } {\cal A}^{\rm R}_{\eta } = 0$, we quickly find that 
\begin{align}
\nonumber 
- \delta _{\Omega } S_{\rm wzw} = \langle D^{\rm NS}_{\eta } \Omega ^{\rm NS} , Q {\cal A}^{\rm NS}_{\eta } + m_{2} ( {\cal A}^{\rm R}_{\eta } , {\cal A}^{\rm R}_{\eta } ) \rangle = 0.   
\end{align}
Therefore, the action $S_{\rm wzw}$ is invariant under $\Omega $-gauge transformations (\ref{delta Omega}). 

\subsection{Unified notation} 

We introduce a notation which is useful to unify the results of NS and R sectors. 
Then, the concept of Ramond number projections proposed in \cite{Erler:2015lya} naturaly appears. 
We say Ramond number of the $k$-product $M_{k}$ is $n$ when number of R imputs of $M_{k}$ minus number of R output of $M_{k}$ equals to $n$. 
The symbol $M_{k}|_{n}$ denotes the $k$-product projected onto Ramond number $n$. 
For example, R number $0$ and $2$ projection of the star product $m_{2}$ are
\begin{align}
&\langle {\rm NS} + {\rm R} , \, m_{2}|_{0}({\rm NS} + {\rm R}, {\rm NS} + {\rm R}) \rangle  = \langle {\rm NS} , \, m_{2}( {\rm NS} , {\rm NS} )\rangle + \langle {\rm R} , \, m_{2} ( {\rm NS} , {\rm R} )+m_{2}( {\rm R} , {\rm NS} )\rangle , 
\no[2mm] \nonumber 
& \hspace{30mm} \langle {\rm NS} + {\rm R} , \, m_{2}|_{2} ({\rm NS} + {\rm R} , {\rm NS} +{\rm R} )\rangle  = \langle {\rm NS} , \,  m_{2} ( {\rm R} , {\rm R} )\rangle . 
\end{align}
It is helpful to specify whether the (output) state ${\cal A}$ is NS or R. 
We write ${\cal A}|^{\rm NS}$ for the NS (output) state and ${\cal A}|^{\rm R}$ for the R (output) state. 
For example, for the sum of NS and R states 
\begin{align}
\nonumber 
( {\rm NS} + {\rm R} )|^{\rm NS} = {\rm NS } , \hspace{10mm } ({\rm NS} + {\rm R})|^{\rm R} = {\rm R} .
\end{align}
Then, we can write as follows: 
\begin{align}
& m_{2}({\rm NS} + {\rm R}, {\rm NS} + {\rm R} ) \big{|}^{\rm NS}_{0} =  m_{2}( {\rm NS} , {\rm NS} ), 
\hspace{5mm} m_{2} ({\rm NS} + {\rm R}, {\rm NS} + {\rm R} ) \big{|}^{\rm R}_{0} = \Ld {\rm NS} , {\rm R} \Rd _{\ast } , 
\no[2mm] \nonumber 
& \hspace{10mm}  m_{2} ({\rm NS} + {\rm R} , {\rm NS} +{\rm R} ) \big{|}^{\rm NS}_{2} = m_{2} ( {\rm R} , {\rm R} ) , \hspace{5mm} m_{2} ({\rm NS} + {\rm R} , {\rm NS} +{\rm R} ) \big{|}^{\rm R}_{2} = 0 . 
\end{align}

\niu{Pure-gauge-like fields and associated fields}

We can introduce a pure-gauge-like (functional) field including both NS and R sectors 
\begin{align}
{\cal A}_{\eta } [\varphi ] \equiv {\cal A}^{\rm NS}_{\eta } [\varphi ] + {\cal A}^{\rm R}_{\eta } [\varphi ] 
\end{align}
such that ${\cal A}_{\eta } = \mathcal{A}_{\eta } [ \varphi ]$ satisfies 
\begin{align}
D_{\eta } {\cal A}_{\eta } \equiv \eta {\cal A}_{\eta } - m_{2}|_{0} ( {\cal A}_{\eta } , {\cal A}_{\eta } ) = 0 . 
\end{align}
In section 3.2, we will explain that $D_{\eta } B \equiv \eta - m_{2}|_{0} (A_{\eta }, B) -(-)^{BA_{\eta }} m_{2}|_{0} (B, A_{\eta })$ is naturally induced from the $A_{\infty }$ products of the WZW-like action: $\eta - m_{2}|_{0}$, which is a dual of $Q + m_{2}|_{2}$. 
Note that NS and R out-puts of $D_{\eta } {\cal A}_{\eta } = 0$ give 
\begin{align} 
\nonumber 
{\rm NS} \, &: \hspace{5mm}  \big( D_{\eta } {\cal A}_{\eta } \big) \big{|} ^{\rm NS} \equiv \eta {\cal A}^{\rm NS}_{\eta } - m_{2} \big( {\cal A}^{\rm NS}_{\eta } , {\cal A}^{\rm NS}_{\eta } \big) = 0 , 
\\ \nonumber 
{\rm R} \, &: \hspace{5mm}   \big( D_{\eta } {\cal A}_{\eta } \big) \big{|} ^{\rm R} \equiv \eta {\cal A}^{\rm R}_{\eta } - \Ld {\cal A}^{\rm NS}_{\eta } , {\cal A}^{\rm R}_{\eta } \Rd _{\ast } = 0 , 
\end{align}
which are just the defining equations of NS and R pure-gauge-like fields (\ref{NS pure-gauge-like equation}) and (\ref{R pure-gauge-like equation}) respectively\footnote{The difference between $D_{\eta}$ and $D^{\rm NS}_{\eta }$ is whether it includes the R-number projection on $m_{2}$ or not. 
While it seems to be trivial for associative open string field theory, it would be highly nontrivial for closed string field theory or generic (nonassociative) open string field theory: 
R-number projections on $\{ m_{n} \} _{n=2}^{\infty } $ should be cralified.}. 
Similarly, we can also define an associated field of $d$ including both sector 
\begin{align}
{\cal A}_{d} [\varphi ] \equiv {\cal A}^{\rm NS}_{d} [\varphi ]  + {\cal A}^{\rm R}_{d} [\varphi ] 
\end{align}
such that ${\cal A}_{d}=\mathcal{A}_{d} [\varphi ]$ satisfies 
\begin{align}
(-)^{d} d \, {\cal A}_{\eta } = D_{\eta } {\cal A}_{d} , 
\end{align}
whose NS out-put $((-)^{d} d {\cal A}_{\eta } = D_{\eta } {\cal A}_{d})|^{\rm NS}$ and R out-put $((-)^{d} d {\cal A}_{\eta } = D_{\eta } {\cal A}_{d})|^{\rm R}$ just provide the defining equations of NS and R pure-gauge-like (functional) fields (\ref{NS pure-gauge-like equation}) and (\ref{R pure-gauge-like equation}) respectively 
\begin{align} 
\nonumber 
{\rm NS} \, &: \hspace{5mm}  (-)^{d} d {\cal A}^{\rm NS}_{\eta } = \eta {\cal A}^{\rm NS}_{d} - \ld {\cal A}^{\rm NS}_{\eta } , {\cal A}^{\rm NS}_{d} \rd _{\ast } , 
\\ \nonumber 
{\rm R} \, &: \hspace{5mm}  (-)^{d} d {\cal A}^{\rm R}_{\eta } = \eta {\cal A}^{\rm R}_{d} - \ld {\cal A}^{\rm NS}_{\eta } , {\cal A}^{\rm R}_{d} \rd _{\ast } - \ld {\cal A}^{\rm R}_{\eta } , {\cal A}^{\rm NS}_{d} \rd _{\ast } .  
\end{align}
Note that in this case, we should take $d$ such that $\ld d, \eta - m_{2}|_{0} \rd $: for example, $d= \partial _{t}$, $\delta $, and $Q+m_{2}|_{2}$. 
See section 3.2 for details and for a quick proof of $\ld \eta - m_{2}|_{0} , Q + m_{2}|_{2} \rd = 0$. 

\niu{Action and equations of motion} 

In this notation, our Wess-Zumino-Witten-like complete action is given by 
\begin{align}
\label{WZW-like complete action}
S_{\rm wzw} [\varphi ] = - \int_{0}^{1} dt \, \langle \mathcal{A}^{\ast }_{t}  \, , \, Q \mathcal{A}_{\eta } + m_{2}|_{2} ( \mathcal{A}_{\eta } , \mathcal{A}_{\eta } ) \rangle , 
\end{align} 
where the associated (functional) field ${\cal A}^{\ast }_{t} = \mathcal{A}^{\ast }_{t} [\varphi ]$ is defined by 
\begin{align}
\mathcal{A}^{\ast }_{t} \equiv {\cal A}^{\rm NS}_{t} + \xi Y \partial _{t} ( P_{\eta } {\cal A}^{\rm R}_{\eta } ) , 
\end{align} 
whose role is explained in section 2.4. 
Note that the projection onto Ramond number $2$ states implies $m_{2}|_{2} ({\cal A}_{\eta } , {\cal A}_{\eta }) = m_{2} ({\cal A}^{\rm R}_{\eta } , {\cal A}^{\rm R}_{\eta } )$ for ${\cal A}^{\rm NS}_{t}$ and $m_{2}|_{2} ( {\cal A}_{\eta } , {\cal A}_{\eta } ) =  0$ for $\xi Y \partial _{t} ( P_{\eta } {\cal A}^{\rm R}_{\eta })$. 
Then, the variation of the action becomes $\delta S_{\rm wzw} = - \langle {\cal A}^{\ast }_{\delta } \, , \, Q {\cal A}_{\eta } + m_{2}|_{2} ({\cal A}_{\eta } , {\cal A}_{\eta } ) \rangle $ with ${\cal A}^{\ast }_{\delta } \equiv {\cal A}^{\rm NS}_{\delta } + \xi Y \delta ( P_{\eta } {\cal A}^{\rm R}_{\eta })$ and the equations of motion is given by 
\begin{align}
Q \mathcal{A}_{\eta } [\varphi ] + m_{2}|_{2} ( \mathcal{A}_{\eta } [\varphi ] , \mathcal{A}_{\eta } [\varphi ] ) = 0 , 
\end{align}
which reproduces NS and R equations of motion (\ref{NS EOM}) and (\ref{R EOM}) by NS and R out-puts projections respectively. 
When we consider another parametrization of the action and its relation to the parametrization given in section 1.1, this notation would be useful.

\subsection{Single functional form}

As we showed, two or more types of functional fields $\mathcal{A}_{\eta }=\mathcal{A}_{\eta } [\varphi ]$, $\mathcal{A}_{t}=\mathcal{A}_{t}[\varphi ]$ appear in the WZW-like action. 
Their algebraic relations make computations easy, but, at the same time, give constraints on these functional fields: 
The existence of many types of (functional) fields satisfying constraint equations would complicate its gauge fixing problem. 
In the rest of this section, we show that one can rewrite the WZW-like action into a form which consists of a single functional field $\mathcal{A}_{\eta }=\mathcal{A}_{\eta }[\varphi ]$ and elementary operators. 

\vspace{2mm} 

We notice that in the first line of (\ref{Original KO}), while the R term consists of a single functional field $\mathcal{A}^{\rm R}_{\eta }$ and operators $\{ Q, \eta , \xi , Y \} $, the NS term includes not only $\mathcal{A}^{\rm NS}_{\eta }$ but $\mathcal{A}^{\rm NS}_{t}$. 
We thus prove that the NS action $S_{\rm NS}$ has a single functional form. 
Recall that the operator $F \xi $ is BPZ even and the derivation $D_{\eta }$ is BPZ odd because $\eta $ is BPZ odd and $\xi $ is given by $P_{\eta } = \eta \xi $. 
Since $D_{\eta } F \xi + F\xi D_{\eta } = 1$ and $D_{\eta } F \xi D_{\eta } = D_{\eta }$ hold, the relation $Q \mathcal{A}^{\rm NS}_{\eta } = - D_{\eta } \mathcal{A}^{\rm NS}_{Q}$ and $(D_{\eta })^{2} = 0$ yield 
\begin{align}
\langle \mathcal{A}^{\rm NS}_{t} , Q \mathcal{A}^{\rm NS}_{\eta }  \rangle 
& = \langle ( F \xi D_{\eta } + D_{\eta } F \xi ) \mathcal{A}^{\rm NS}_{t} , Q \mathcal{A}^{\rm NS}_{\eta }  \rangle  
= \langle F \xi ( D_{\eta } \mathcal{A}^{\rm NS}_{t} ) , Q \mathcal{A}^{\rm NS}_{\eta }  \rangle 
\no \nonumber
& = \langle F \xi ( \partial _{t}  \mathcal{A}^{\rm NS}_{\eta } ) , Q \mathcal{A}^{\rm NS}_{\eta }  \rangle  
= -  \langle \partial _{t}  \mathcal{A}^{\rm NS}_{\eta }  , F \xi ( Q \mathcal{A}^{\rm NS}_{\eta } ) \rangle . 
\end{align}
Here, we used not $D^{\rm NS}_{\eta }$ but $D_{\eta }$: Both work well. 
One can check that $\ld D_{\eta } , F\xi \rd = 1$, $\xi F = \xi$, and so on in the completely same manner as those of $\ld D^{\rm NS}_{\eta }, F \xi \rd = 1$ and so on, if we use the following definition of $F$, which is independent\footnote{We would like to emphasise that one can define this linear operator $F$ provided that the $A_{\infty }$ products is identified: 
In this paper, it is $\eta - m_{2}|_{0}$, which is the dual of $Q + m_{2}|_{2}$.
$D_{\eta }$ is the $1$-slot shifted product which is generated by the Maurer-Cartan element $\mathcal{A}_{\eta }$ of the $A_{\infty }$ products.
See section 3.2.} of our choice of dynamical string fields, 
\begin{align}
F \equiv \sum_{n=0}^{\infty } \big[ \xi ( \eta - D_{\eta } ) \big] ^{n} . 
\end{align} 
This $F$ consists of $\mathcal{A}_{\eta }$, $\xi $, $\eta $, and $\eta - m_{2}|_{0}$. 
We can rewrite the action into a form which consists of single WZW-like functional $\mathcal{A}^{\rm NS}_{\eta }$ and operators $Q$, $\eta $, $\xi $, $\partial _{t}$, $\eta - m_{2}|_{0}$ as follows 
\begin{align}
S^{\rm NS} [\varphi ] & = - \int_{0}^{1} dt \, \langle \mathcal{A}^{\rm NS}_{t} , Q \mathcal{A}^{\rm NS}_{\eta }  \rangle 
= \int_{0}^{1} dt \, \langle \partial _{t}  \mathcal{A}^{\rm NS}_{\eta }  , F \xi  Q \mathcal{A}^{\rm NS}_{\eta }  \rangle 
\no 
& = \int_{0}^{1} dt \, \sum_{n=0}^{\infty } \, \langle \partial _{t} \mathcal{A}^{\rm NS}_{\eta } , \xi \big[ (\eta - D_{\eta}) \xi \big] ^{n} Q \mathcal{A}^{\rm NS}_{\eta } \rangle .  
\end{align}

One can quickly find that this single functional form $S_{\rm NS} = \int_{0}^{1} dt \langle \partial _{t} \mathcal{A}^{\rm NS}_{\eta } , F \xi Q \mathcal{A}^{\rm NS}_{\eta } \rangle $ also has topological $t$-dependence, $\delta S_{\rm NS} = \langle \delta \mathcal{A}^{\rm NS}_{\eta }, F \xi Q \mathcal{A}^{\rm NS}_{\eta }\rangle $. 
Note that from the above definition of $F$, the following relation holds: 
\begin{align}
\Ld d , F \xi \Rd & = - F \xi \Ld d , D_{\eta } \Rd F \xi + \Ld D_{\eta } , F \xi d F \xi \Rd 
\no 
& = F \xi \Ld d A_{\eta } , \hspace{3mm} \Rd _{\ast } F \xi + \Ld D_{\eta } , F \xi d F \xi \Rd . 
\end{align}
It implies that we can convert the graded commutator of operators $\Ld d , F \xi \Rd $ into that of the star product $F \xi \Ld d A_{\eta } , \,\,\, \Rd _{\ast } F \xi $ in the inner product of two $D_{\eta }$-exact states. 
Using this relation, one can quickly find that as well as the original WZW-like one, this single functional form also has the topological $t$-dependence. 
In the following computations, we write $\mathcal{A}_{\eta }$ for $\mathcal{A}^{\rm NS}_{\eta }$ because both computations are same and our goal is to obtain a single form of the complete action (\ref{WZW-like complete action}). 
The variation of the NS action is given by 
\begin{align}
\delta S [\varphi ] & = \int_{0}^{1} dt \, \Big( \langle \partial _{t} ( \delta \mathcal{A}_{\eta } ) , F \xi Q \mathcal{A}_{\eta } \rangle +\langle \partial _{t} \mathcal{A}_{\eta } , \delta ( F \xi Q \mathcal{A}_{\eta } ) \rangle \Big) 
\no \nonumber 
& = \langle \delta \mathcal{A}_{\eta } , F \xi Q \mathcal{A}_{\eta } \rangle + \int_{0}^{1} dt \, \Big[ \langle \partial _{t} \mathcal{A}_{\eta } , \delta ( F \xi Q \mathcal{A}_{\eta } ) \rangle - \langle \delta \mathcal{A}_{\eta } , \partial _{t} ( F \xi Q \mathcal{A}_{\eta } ) \rangle \Big] . 
\end{align}
The integrand of the second term of the second line $\langle \partial _{t} \mathcal{A}_{\eta } , \delta ( F \xi Q \mathcal{A}_{\eta } ) \rangle - \langle \delta \mathcal{A}_{\eta } , \partial _{t} ( F \xi Q \mathcal{A}_{\eta } ) \rangle $, which is the extra term, becomes 
\begin{align} 
& \Big[ \langle \partial _{t} \mathcal{A}_{\eta } , F \xi Q ( \delta \mathcal{A}_{\eta } ) \rangle - \langle \delta \mathcal{A}_{\eta } , F \xi Q ( \partial _{t} \mathcal{A}_{\eta } ) \rangle \Big] 
- \langle \partial _{t} \mathcal{A}_{\eta } ,  F \xi \ld \delta , D_{\eta } \rd F \xi Q \mathcal{A}_{\eta } \rangle + \langle \delta \mathcal{A}_{\eta } , F \xi \ld \partial_{t} , D_{\eta } \rd F \xi Q \mathcal{A}_{\eta }  \rangle
\no 
&\hspace{15mm} = - \langle \delta \mathcal{A}_{\eta } , \ld Q , F \xi \rd \partial _{t} \mathcal{A}_{\eta } \rangle 
+ \langle \partial _{t} \mathcal{A}_{\eta } , \Ld \delta , F \xi \Rd Q \mathcal{A}_{\eta } \rangle - \langle \delta \mathcal{A}_{\eta } , \Ld \partial _{t} , F \xi \Rd Q \mathcal{A}_{\eta } \rangle 
\no 
&\hspace{15mm} = - \langle \delta \mathcal{A}_{\eta } , \ld Q , F \xi \rd \partial _{t} \mathcal{A}_{\eta } \rangle 
- \langle \partial _{t} \mathcal{A}_{\eta } , F \xi \Ld \delta \mathcal{A}_{\eta } , F \xi Q \mathcal{A}_{\eta } \Rd _{\ast } \rangle 
+ \langle \delta \mathcal{A}_{\eta } , F \xi \Ld \partial _{t} \mathcal{A}_{\eta } , F \xi Q \mathcal{A}_{\eta } \Rd _{\ast } \rangle 
\no \nonumber 
&\hspace{15mm} = - \langle \delta \mathcal{A}_{\eta } , \ld Q , F \xi \rd \partial _{t} \mathcal{A}_{\eta } \rangle 
+ \langle \delta \mathcal{A}_{\eta } , F \xi \Ld Q \mathcal{A}_{\eta } , F \xi \partial _{t} \mathcal{A}_{\eta } \Rd _{\ast } \rangle = 0 .
\end{align}
Therefore, we obtain $\delta S = \langle \delta \mathcal{A}_{\eta } , F \xi Q \mathcal{A}_{\eta } \rangle $. 

\niu{Single functional form}

Similarly, we can write the following form of the action using single WZW-like functional $\mathcal{A}_{\eta } = \mathcal{A}^{\rm NS}_{\eta } + \mathcal{A}^{\rm R}_{\eta }$ and operators $Q+m_{2}|_{2}$, $\xi$, $\partial _{t}$, $\eta - m_{2}|_{0}$,   
\begin{align}
\label{Single functional form}
S_{\rm wzw} [\varphi ] = \int_{0}^{1} dt \, \langle \partial _{t}  \mathcal{A}_{\eta } , \,  ( F \xi + \xi Y ) \Big[ Q \mathcal{A}_{\eta } + m_{2}|_{2} ( \mathcal{A}_{\eta } , \mathcal{A}_{\eta } ) \Big] \rangle . 
\end{align}
Note that $\langle {\rm NS} , {\rm R} \rangle = \langle {\rm R} , {\rm NS} \rangle = 0$ and that $\langle A , B \rangle = 0$ when the sum of ghost-and-picture numbers of $A$ and $B$ is not $(2|-1)$. 
Its $t$-dependence is again topological: 
\begin{align}
\delta S_{\rm wzw} [\varphi ] 
& = \langle  \delta \mathcal{A}_{\eta } , \,  ( F \xi + \xi Y ) \big[ Q \mathcal{A}_{\eta } + m_{2}|_{2} ( \mathcal{A}_{\eta } , \mathcal{A}_{\eta } ) \big] \rangle . 
\end{align}
From this form, we find that the associated (functional) field $\mathcal{A}^{\ast }_{t}$ appearing in the action is
\begin{align}
\mathcal{A}^{\ast }_{t} = ( F \xi + Y \xi ) \partial _{t} \mathcal{A}_{\eta } , 
\end{align}
which consists of a single functional field $\mathcal{A}_{\eta }$ and operators. 
It enable us to rewrite the action into a single functional form and reminds us constraints on the state space spanned by $\mathcal{A}_{\eta }$: 
$\eta \mathcal{A}_{\eta } - m_{2}|_{0} ( \mathcal{A}_{\eta } , \mathcal{A}_{\eta } ) = 0$, which yields $d \mathcal{A}_{\eta } = D_{\eta } F \xi ( d \mathcal{A}_{\eta } )$, and $\xi \mathcal{A}^{\rm R}_{\eta} = XY \xi \mathcal{A}^{\rm R}_{\eta }$.

\section{Another parametrization} 

We use the same notation as \cite{Erler:2015lya} in this section. 
Readers who are unfamiliar with $A_{\infty }$ algebras or coalgebraic computations see, for example, \cite{Erler:2015lya,Erler:2015rra, Erler:2015uba, Erler:2015uoa, GM2, Erler:2013xta, Erler:2014eba, Gaberdiel:1997ia} or other mathematical manuscripts\cite{Bar complex, Penkava, Kajiura:2003ax}. 
In the work of \cite{Erler:2015lya}, the on-shell conditions of superstring field theories are proposed. 
For open superstring field theory, it is given by 
\begin{align}
\label{EKS eom}
\pi _{1} \big( {\boldsymbol Q} + {\boldsymbol m}_{2}|_{2} \big) \, \widehat{{\bf G}} \frac{1}{1 - \widetilde{\Psi } } = 0 ,
\end{align}
where $\widetilde{\Psi } = \widetilde{\Psi }^{\rm NS} + \widetilde{\Psi }^{\rm R}$ is an NS plus R string field, $Q$ is the BRST operator, and $m_{2}|_{2}$ denotes the star product $m_{2}$ with R number 2 projection. 
$\pi _{1}$ denotes the projection onto the single state space $\mathcal{H}$ from $T(\mathcal{H}) = \bigoplus _{n} \mathcal{H}^{\otimes n}$. 
Note that NS and R out-puts of (\ref{EKS eom}) are given by 
\begin{align}
\label{EKS NS eom}
{\rm NS} \, & : \hspace{5mm}  Q \, \pi _{1} \widehat{{\bf G}} \frac{1}{1 - \widetilde{\Psi } } \Big{|}^{\rm NS} + m_{2} \Big( \pi _{1} \widehat{{\bf G}} \frac{1}{1 - \widetilde{\Psi } } \Big{|}^{\rm R} , \pi _{1} \widehat{{\bf G}} \frac{1}{1 - \widetilde{\Psi } } \Big{|}^{\rm R} \Big)  = 0 ,
\\ 
\label{EKS R eom}
{\rm R} \, & : \hspace{30mm} Q \, \pi _{1} \widehat{{\bf G}} \frac{1}{1 - \widetilde{\Psi } } \Big{|}^{\rm R} = 0 . 
\end{align}
Note also that in general, the cohomomorphism $\widehat{\bf G}$ is constructed by the path-ordered exponential (with direction) of a gauge product ${\boldsymbol \mu }(t)$, a coderivation, as follows 
\begin{align}
\label{exp} 
\widehat{\bf G} \equiv \mathcal{P} \exp{\Big[  \int_{0}^{t} d t' \, {\boldsymbol \mu } (t') \Big] }  . 
\end{align}

In this paper, we always use $\widehat{\bf G}$ given in \cite{Erler:2015lya}. 
In \cite{Erler:2015lya}, the gauge product ${\boldsymbol \mu }(t)$ consists of R number $0$ projected objects. 
Therefore, $\pi _{1} \widehat{\bf G}$ has at most one Ramond state in its in-puts. 

\subsection{Another parametrization of the WZW-like complete action}

In this section, we define pure-gauge-like and associated fields parametrized by $\widetilde{\Psi }= \widetilde{\Psi }^{\rm NS} + \widetilde{\Psi }^{\rm R}$ and construct a gauge invariant action, whose equations of motion equals to (\ref{EKS eom}), the Ramond equations of motion proposed in \cite{Erler:2015lya}. 
The proofs of required properties are in section 3.2.

\niu{Parametrization inspired by Ramond equations of motion}

We can construct an NS pure-gauge-like (functional) field $\widetilde{A}^{\rm NS}_{\eta } = \widetilde{A}^{\rm NS}_{\eta }[\widetilde{\Psi } ]$ by 
\begin{align} 
\label{NS pure-gauge-like tilde}
\widetilde{A}^{\rm NS}_{\eta } \equiv \pi _{1} \widehat{\bf G} \, \frac{1}{1- \widetilde{\Psi }} \Big{|}^{\rm NS} 
= \pi _{1} \widehat{\bf G} \, \frac{1}{1- \widetilde{\Psi }^{\rm NS}} , 
\end{align}
and a R pure-gauge-like (functional) field $\widetilde{A}^{\rm R}_{\eta } = \widetilde{A}^{\rm NS}_{\eta } [\widetilde{\Psi } ]$ by
\begin{align}
\label{R pure-gauge-like tilde}
\widetilde{A}^{\rm R}_{\eta } \equiv \pi _{1} \widehat{\bf G} \, \frac{1}{1- \widetilde{\Psi }} \Big{|}^{\rm R}
= \pi _{1} \widehat{\bf G} \, \Big( \frac{1}{1-\widetilde{\Psi }^{\rm NS}} \otimes \widetilde{\Psi }^{\rm R} \otimes \frac{1}{1- \widetilde{\Psi }^{\rm NS}} \Big) . 
\end{align}
These pure-gauge-like fields are parametrized by NS and R string field $\widetilde{\Psi } = \widetilde{\Psi }^{\rm NS} + \widetilde{\Psi }^{\rm R}$. 
While the NS string field $\widetilde{\Psi }$ is a Grassmann odd and ghost-and-picture number $(1|-1)$ state in the small Hilbert space $\mathcal{H}_{\rm S}$, the R string field $\widetilde{\Psi }^{\rm R}$ is a Grassmann odd and ghost-and-picture number $(1|-\frac{1}{2})$ state in the restricted small Hilbert space $\mathcal{H}_{R}$. 
Hence, $\widetilde{\Psi }^{\rm NS} \in \mathcal{H}_{\rm S}$ and $\widetilde{\Psi }^{\rm R} \in \mathcal{H}_{R}$ satisfy 
\begin{align}
\nonumber 
\eta \, \widetilde{\Psi }^{\rm NS} = 0, \hspace{5mm}  \eta \, \widetilde{\Psi }^{\rm R} = 0, \hspace{5mm} XY \widetilde{\Psi }^{\rm R} = \widetilde{\Psi }^{\rm R} . 
\end{align}
Note that $\widetilde{A}^{\rm NS}_{\eta }$ has ghost-and-picture number $(1 | -1)$ and $\widetilde{A}^{\rm R}_{\eta }$ has ghost-and-picture number $(1|-\frac{1}{2} )$ by construction. 
As we will see in section 3.2, one can check that $\widetilde{A}^{\rm NS}_{\eta }$ and $\widetilde{A}^{\rm R}_{\eta }$ satisfy the defining properties of pure-gauge-like fields: 
\begin{align}
\label{EKS NS pure-gauge-like eq.}
{\rm NS} \, & : \hspace{13mm} \eta \widetilde{A}^{\rm NS}_{\eta } - m_{2}|_{0} ( \widetilde{A}^{\rm NS}_{\eta } , \widetilde{A}^{\rm NS}_{\eta } ) = 0 ,
\\[2mm] 
\label{EKS R pure-gauge-like eq.}
{\rm R} \, & : \hspace{5mm} \eta \widetilde{A}^{\rm R}_{\eta } - m_{2}|_{0} ( \widetilde{A}^{\rm NS}_{\eta } , \widetilde{A}^{\rm R}_{\eta } ) - m_{2}|_{0} ( \widetilde{A}^{\rm R}_{\eta } , \widetilde{A}^{\rm NS}_{\eta } ) = 0 . 
\end{align}
Let $d$ be a derivation operator commuting with $\eta $. 
For example, one can take $d = Q , \partial _{t} , \delta $. 
Then, with these pure-gauge-like fields $\widetilde{A}^{\rm NS}_{\eta }[\widetilde{\Psi } ]$, $\widetilde{A}^{\rm R}_{\eta }[\widetilde{\Psi } ]$ parametrized by small Hilbert space string fields $\widetilde{\Psi }$, an NS associated (functional) field $\widetilde{A}^{\rm NS}_{d} = \widetilde{A}^{\rm NS}_{d} [\widetilde{\Psi } ]$ defined by
\begin{align} 
\label{NS associated tilde}
\widetilde{A}^{\rm NS}_{d} & \equiv \pi _{1}  \widehat{\bf G} \Big( {\boldsymbol \xi }_{{\boldsymbol d}} \frac{1}{1- \widetilde{\Psi } } \Big) \Big{|}^{\rm NS}  
\end{align} 
and a R associated (functional) field $\widetilde{A}^{\rm R}_{d} = \widetilde{A}^{\rm R}_{d} [\widetilde{\Psi } ]$ defined by 
\begin{align} 
\label{R associated tilde}
\widetilde{A}^{\rm R}_{d} & \equiv \pi _{1}  \widehat{\bf G} \Big( {\boldsymbol \xi }_{{\boldsymbol d}} \frac{1}{1- \widetilde{\Psi } } \Big) \Big{|}^{\rm R} 
\end{align}
satisfy the defining properties of associated fields: 
\begin{align}
\label{NS associated eq. tilde}
(-)^{d} d  \, \widetilde{A}^{\rm NS}_{\eta } & =  \eta  \widetilde{A}^{\rm NS}_{d}  - \Ld \widetilde{A}^{\rm NS}_{\eta } , \widetilde{A}^{\rm NS}_{d} \Rd _{\ast } , 
\\[2mm] \label{R associated eq. tilde}
(-)^{d} d \, \widetilde{A}^{\rm R}_{\eta } & = \eta \widetilde{A}^{\rm R}_{d} - \Ld \widetilde{A}^{\rm NS}_{\eta } , \widetilde{A}^{\rm R}_{d} \Rd _{\ast } - \Ld \widetilde{A}^{\rm R}_{\eta } , \widetilde{A}^{\rm NS}_{d} \Rd _{\ast } , 
\end{align} 
which we prove in section 3.2. 
Once the defining properties (\ref{EKS NS pure-gauge-like eq.}), (\ref{EKS R pure-gauge-like eq.}), (\ref{NS associated eq. tilde}), and (\ref{R associated eq. tilde}) are proved using pure-gauge-like fields $\widetilde{A}^{\rm NS}_{\eta }$, $\widetilde{A}^{\rm R}_{\eta }$ defined by (\ref{NS pure-gauge-like tilde}), (\ref{R pure-gauge-like tilde}) and associated fields $\widetilde{A}^{\rm NS}_{d}$, $\widetilde{A}^{\rm R}_{d}$ defined by (\ref{NS associated tilde}), (\ref{R associated tilde}), we can construct a gauge invariant action on the basis of Wess-Zumino-Witten-like framework proposed in section 2. 

\niu{Consistency with the XY-projection}

To apply our WZW-like framework, we need the $XY$-projection invariance of $P_{\eta } \widetilde{A}^{\rm R}_{\eta }$ 
\begin{align}
\label{XY projection tilde}
X Y ( P_{\eta } \widetilde{A}^{\rm R}_{\eta } ) = P_{\eta } \widetilde{A}^{\rm R}_{\eta } . 
\end{align}
Unfortunately, for any choice of cohomomorphism $\widehat{\bf G}$, the R pure-gauge-like field $\widetilde{A}^{\rm R}_{\eta }$ defined by (\ref{R pure-gauge-like tilde}) does not satisfy this property. 
Note, however, that if we can take $\widehat{\bf G}$ satisfying 
\begin{align} 
\label{xi G = xi}
{\boldsymbol \xi } \, \widehat{\bf G} = {\boldsymbol \xi }  
\end{align}
on the R out-put state, then the R pure-gauge-like (functional) field $\widetilde{A}^{\rm R}_{\eta }[\widetilde{\Psi } ]$ defined by (\ref{R pure-gauge-like tilde}) automatically satisfy (\ref{XY projection tilde}) because 
\begin{align}
X Y ( P_{\eta } \widetilde{A}^{\rm R}_{\eta } ) & = X Y P_{\eta } \, \pi _{1} \widehat{\bf G} \frac{1}{1 - \widetilde{\Psi } } \Big{|}^{\rm R} =  X Y \eta \, \pi _{1} {\boldsymbol \xi } \widehat{\bf G} \frac{1}{1 - \widetilde{\Psi } } \Big{|}^{\rm R} 
\no \nonumber 
& =   X Y P_{\eta } \, \pi _{1} \frac{1}{1 - \widetilde{\Psi } } \Big{|}^{\rm R} 
= X Y P_{\eta } \widetilde{\Psi }^{\rm R}  
= P_{\eta } \widetilde{\Psi }^{\rm R} = P_{\eta } \widetilde{A}^{\rm R}_{\eta } , 
\end{align}
with $X Y \widetilde{\Psi }^{\rm R} = \widetilde{\Psi }^{\rm R}$ and $P_{\eta } = \eta \xi$ where $\xi = \xi _{0}$ for NS states and $\xi = \Xi $ for R states. 
Recall that $\widehat{\bf G}$ is constructed by the path-ordered exponential of a gauge product ${\boldsymbol \mu }(t)$ as (\ref{exp}). 
When we take this gauge product as $\xi $-exact one ${\boldsymbol \mu } (t) \equiv {\boldsymbol \xi } {\boldsymbol M} (t)$, the cohomomorphism $\widehat{\bf G}$ is given by 
\begin{align}
\nonumber 
\widehat{\bf G} \equiv \mathcal{P} e^{\int dt \, {\boldsymbol \xi } {\boldsymbol M} (t) } = \mathbb{I} + {\boldsymbol \xi } \Big( M_{2} + \frac{1}{2}M_{3} + \frac{1}{2} M_{2} \xi M_{2} + \dots \Big) , 
\end{align}
and it satisfies (\ref{xi G = xi}). 
This $\xi $-exact choice of the gauge product is always possible by using ambiguities of the construction of (intermediate) gauge products ${\boldsymbol \mu }(t)$ or setting the initial condition of the defining equations of the $A_{\infty }$ products in \cite{Erler:2015lya}. 
Note that although a naive choice of $\xi $-exact gauge products ${\boldsymbol \mu } = {\boldsymbol \xi } {\boldsymbol M}$ breaks the ciclic property of the $A_{\infty }$ products $\widetilde{\boldsymbol M} = \widehat{\bf G}^{-1} \, ({\boldsymbol Q} + {\boldsymbol m}_{2}|_{2} ) \, \widehat{\bf G}$, it is no problem in our Wess-Zumino-Witten-like framework: 
$Q+m_{2}|_{2}$ and $\eta -m_{2}|_{0}$ work well in the inner product with appropriate states, namely, in the action. 

\vspace{2mm} 

We would like to emphasize that it does not necessitate the ciclic property of $\widehat{\bf G}$ or $\widetilde{\boldsymbol M}$ to construct the Wess-Zumino-Witten-like complete action. 
We need the ciclic property of $D_{\eta }$ only (namely, its BPZ oddness), which holds for any choice of $\widehat{\bf G}$. 
Hence, we can always impose (\ref{XY projection tilde}) in a consistent way with the definitions of pure-gauge-like fields (\ref{NS pure-gauge-like tilde}), (\ref{R pure-gauge-like tilde}). 

\niu{Action and gauge invariance}

Utilizing pure-gauge-like and associated fields satisfying (\ref{EKS NS pure-gauge-like eq.}), (\ref{EKS R pure-gauge-like eq.}), (\ref{NS associated eq. tilde}), (\ref{R associated eq. tilde}), and (\ref{XY projection tilde}), we construct the Wess-Zumino-Witten-like complete action by $\widetilde{S} [\widetilde{\Psi }] \equiv S_{\rm wzw} [\widetilde{\Psi }] $, 
\begin{align}
\label{S tilde}
\widetilde{S} [\widetilde{\Psi }] = \int_{0}^{1} dt \Big( \langle \xi Y \partial _{t} ( P_{\eta } \widetilde{A}^{\rm R}_{\eta } ) , Q \widetilde{A}^{\rm R}_{\eta } \rangle + \langle \widetilde{A}^{\rm NS}_{t} , Q \widetilde{A}^{\rm NS}_{\eta } + m_{2} ( \widetilde{A}^{\rm R}_{\eta } , \widetilde{A}^{\rm R}_{\eta } ) \rangle \Big) , 
\end{align} 
which is parametrized by small Hilbert space string fields $\widetilde{\Psi } = \widetilde{\Psi }^{\rm NS} + \widetilde{\Psi }^{\rm R}$. 
As we found in section 2, the variation of the action is given by 
\begin{align} 
\label{delta S tilde} 
\delta \widetilde{S} [\widetilde{\Psi }]  = \langle \xi Y \delta ( P_{\eta } \widetilde{A}^{\rm R}_{\eta } ) , Q \widetilde{A}^{\rm R}_{\eta } \rangle + \langle \widetilde{A}^{\rm NS}_{\delta } , Q \widetilde{A}^{\rm NS}_{\eta } + m_{2} ( \widetilde{A}^{\rm R}_{\eta } , \widetilde{A}^{\rm R}_{\eta } ) \rangle , 
\end{align} 
and the action is invariant under two types of gauge transformations: 
the gauge transformations parametrized by $\Lambda = \Lambda ^{\rm NS} + \Lambda ^{\rm R}$  
\begin{align}
\nonumber 
\widetilde{A}^{\rm NS}_{\delta _{\Lambda }} & = Q \Lambda ^{\rm NS} + \Ld \widetilde{A}^{\rm R}_{\eta } , \Lambda ^{\rm R} \Rd _{\ast } , 
\\ \nonumber 
\delta _{\Lambda } ( P_{\eta } \widetilde{A}^{\rm R}_{\eta } ) & = - P_{\eta } Q \Big( \eta \Lambda ^{\rm R} - \Ld \widetilde{A}^{\rm NS}_{\eta } , \Lambda ^{\rm R} \Rd _{\ast } - \Ld \widetilde{A}^{\rm R}_{\eta } , \Lambda ^{\rm NS} \Rd _{\ast } \Big) , 
\end{align}
and the gauge transformations parametrized by $\Omega = \Omega ^{\rm NS}$ 
\begin{align}
\nonumber 
\widetilde{A}^{\rm NS}_{\delta _{\Omega } }  = \eta \Omega ^{\rm NS} - \Ld \widetilde{A}^{\rm NS}_{\eta } , \Omega ^{\rm NS} \Rd _{\ast } , 
\hspace{5mm}  
\delta _{\Omega } ( P_{\eta } \widetilde{A}^{\rm R}_{\eta } ) = 0 . 
\end{align} 
Here NS, R, and NS gauge parameter fields $\Lambda ^{NS}$, $\Lambda ^{\rm R}$, and $\Omega ^{\rm NS}$ have ghost-and-picture number $(-1|0)$, $(-1|\frac{1}{2})$, and $(-1|1)$, respectively, and these fields all belong to the large Hilbert space. 
Note, however, that the gauge transformation $\delta _{\Lambda } (P_{\eta } \widetilde{A}^{\rm R}_{\eta })$ has to be in the restricted small Hilbert space $\mathcal{H}_{R}$. 

\vspace{2mm} 

\niu{Equations of motion}

Since the action $\widetilde{S}$ has topological $t$-dependence and its variation is given by (\ref{delta S tilde}), we obtain the equations of motion 
\begin{align}
\label{NS eom tilde}
{\rm NS} \, : & \hspace{5mm}   Q \widetilde{A}^{\rm NS}_{\eta } [\widetilde{\Psi }] + m_{2} ( \widetilde{A}^{\rm R}_{\eta }  [\widetilde{\Psi }] , \widetilde{A}^{\rm R}_{\eta } [\widetilde{\Psi }]  )  = \pi _{1} \big( {\boldsymbol Q} + {\boldsymbol m}_{2}|_{2} \big) \, \widehat{{\bf G}} \frac{1}{1 - \widetilde{\Psi } } \Big{|}^{\rm NS} = 0 ,
\\ \label{R eom tilde}
{\rm R} \, : & \hspace{10mm} P_{\eta } \big( Q \widetilde{A}^{\rm R}_{\eta } [\widetilde{\Psi }] \big) = P_{\eta } \Big( \pi _{1} \big( {\boldsymbol Q} + {\boldsymbol m}_{2}|_{2} \big) \, \widehat{{\bf G}} \frac{1}{1 - \widetilde{\Psi } } \Big) \Big{|}^{\rm R}  = 0 , 
\end{align} 
which is equivalent to (\ref{EKS eom}). 
While the NS out-put of the equations of motion (\ref{NS eom tilde}) is the same as (\ref{EKS NS eom}), the R out-put (\ref{R eom tilde}) is equal to the small Hilbert space component of (\ref{EKS R eom}). 
Note that $P_{\xi } ( Q \widetilde{A}^{\rm R}_{\eta } )$ can not be determined from the action because it vanishes in the inner product, and it does not affect the value of the action. 
We thus set $P_{\xi } ( Q \widetilde{A}^{\rm R}_{\eta } ) = 0$ and obtain (\ref{EKS R eom}). 

\niu{Kinetic term}

It is interesting to compare kinetic terms of (\ref{S tilde}) and (\ref{S KO}). 
In the present parametrization of (\ref{S tilde}), the kinetic term of $\widetilde{S}$ is given by 
\begin{align}
\nonumber 
- \frac{1}{2} \langle \xi \widetilde{\Psi }^{\rm NS} , Q \widetilde{\Psi }^{\rm NS} \rangle - \frac{1}{2} \langle \xi Y \widetilde{\Psi }^{\rm R} , Q \widetilde{\Psi }^{\rm R} \rangle . 
\end{align}
Note that the Ramond kinetic term is just equal to that of Kunitomo-Okawa's action. 
Similarly, we quickly check that the NS kinetic term is equivalent to that of Kunitomo-Okawa's action with the (trivial emmbeding) condition $\widetilde{\Psi }^{\rm NS} = \eta \Phi ^{\rm NS}$ or the (linear partial gauge fixing) condition $\Phi ^{\rm NS} = \xi \widetilde{\Psi }^{\rm NS}$. 
Therefore, the kinetic term of $\widetilde{S}  [\widetilde{\Psi }] $ has the same spectrum as that of \cite{Kunitomo:2015usa}.

\subsection{WZW-like relation from $A_{\infty }$ and $\eta $-exactness}

\niu{Dual $A_{\infty }$-products and Derivation properties}

Let ${\boldsymbol \eta }$ be the coderivation constructed from $\eta $, which is nilpotent ${\boldsymbol \eta }^{2} = 0$, and let ${\boldsymbol a }$ be a nilpotent coderivation satisfying ${\boldsymbol a} {\boldsymbol \eta } = -(-)^{{\boldsymbol a} {\boldsymbol \eta }} {\boldsymbol \eta } {\boldsymbol a}$ and ${\boldsymbol a}^{2} = 0$. 
Then, we assume that $\widehat{\bf G}^{-1} : (\mathcal{H} , {\boldsymbol a}) \rightarrow (\mathcal{H}_{\rm S} , {\boldsymbol D}_{\boldsymbol a})$ is an $A_{\infty }$-morphism, where $\mathcal{H}$ is the large Hilbert space, $\mathcal{H}_{\rm S}$ is the small Hilbert space, and ${\boldsymbol D}_{\boldsymbol a} \equiv \widehat{\bf G}^{-1} {\boldsymbol a} \, \widehat{\bf G}$. 
Note that ${\boldsymbol D}_{\boldsymbol a}$ is nilpotent: ${\boldsymbol D}_{\boldsymbol a}^{2} = ( \widehat{\bf G}^{-1} {\boldsymbol a} \, \widehat{\bf G} ) ( \widehat{\bf G}^{-1} {\boldsymbol a} \, \widehat{\bf G} ) = \widehat{\bf G}^{-1} {\boldsymbol a}^{2} \, \widehat{\bf G} = 0$. 
For example, one can use ${\boldsymbol Q}$, ${\boldsymbol Q}+{\boldsymbol m}_{2}|_{2}$, and so on for ${\boldsymbol a}$, and various $\widehat{\bf G}$ appearing in \cite{Erler:2013xta, Erler:2014eba, Erler:2015lya} for $\widehat{\bf G}$. 
Suppose that the coderivation ${\boldsymbol D}_{\boldsymbol a}$ also commutes with ${\boldsymbol \eta }$, which means 
\begin{align} 
\label{[D_a,eta]=0}
( {\boldsymbol D}_{\boldsymbol a} )^{2} = 0 , \hspace{5mm} \Ld {\boldsymbol D}_{\boldsymbol a} , {\boldsymbol \eta } \Rd = 0. 
\end{align} 
Then, we can introduce a dual $A_{\infty }$-products ${\boldsymbol D}^{\boldsymbol \eta }$ defined by 
\begin{align} 
{\boldsymbol D}^{\boldsymbol \eta } \equiv \widehat{\bf G} \, {\boldsymbol \eta } \, \widehat{\bf G}^{-1} . 
\end{align}
Note that the pair of nilpotent maps $( {\boldsymbol D}^{\boldsymbol \eta } ,{\boldsymbol a})$ have the same properties as $( {\boldsymbol D}_{\boldsymbol a} , {\boldsymbol \eta })$: 
\begin{align} 
\label{[D^eta,a]=0}
( {\boldsymbol D}^{\boldsymbol \eta } ) ^{2} = 0 , \hspace{5mm} \Ld {\boldsymbol D}^{\boldsymbol \eta } , {\boldsymbol a} \Rd = 0 . 
\end{align} 
We can quickly find when the $A_{\infty }$-products ${\boldsymbol D}_{\boldsymbol a}$ commutes with the coderivation ${\boldsymbol \eta }$ as (\ref{[D_a,eta]=0}), its dual $A_{\infty }$-product ${\boldsymbol D}^{\boldsymbol \eta }$ and coderivation ${\boldsymbol a}$ also satisfies (\ref{[D^eta,a]=0}) as follows 
\begin{align} 
{\boldsymbol a} {\boldsymbol D}^{\boldsymbol \eta } &= \big( \widehat{\bf G} \, \widehat{\bf G}^{-1} \big) \, {\boldsymbol a} \, \big( \widehat{\bf G} \, {\boldsymbol \eta } \, \widehat{\bf G}^{-1} \big) = \widehat{\bf G} \, {\boldsymbol D}_{\boldsymbol a} \,  {\boldsymbol \eta } \, \widehat{\bf G}^{-1} 
\no \nonumber 
& = (-)^{{\boldsymbol a}{\boldsymbol \eta }} \widehat{\bf G} \, {\boldsymbol \eta } \, {\boldsymbol D}_{\boldsymbol a}  \, \widehat{\bf G}^{-1} 
= (-)^{{\boldsymbol a}{\boldsymbol \eta }} \, \widehat{\bf G} \, {\boldsymbol \eta } \, \widehat{\bf G}^{-1} \, {\boldsymbol a} \, \widehat{\bf G} \, \widehat{\bf G}^{-1} 
= (-)^{{\boldsymbol a}{\boldsymbol \eta }} {\boldsymbol D}^{\boldsymbol \eta } \, {\boldsymbol a} .
\end{align}

In this paper,  as these coderication ${\boldsymbol a}$ and $A_{\infty }$-morphism $\widehat{\bf G}$, we always use ${\boldsymbol a} \equiv {\boldsymbol Q} + {\boldsymbol m}_{2}|_{2}$ and $\widehat{\bf G}$ introduced in (\ref{xi G = xi}), namely, a gauge product $\widehat{\bf G}$ given by \cite{Erler:2015lya} with the choice satisfying ${\boldsymbol \xi} \, \widehat{\bf G} = {\boldsymbol \xi }$. 
Therefore, the dual $A_{\infty }$ products is always given by  
\begin{align} 
\label{D^eta}
{\boldsymbol D}^{\boldsymbol \eta } = {\boldsymbol \eta } - {\boldsymbol m}_{2}|_{0} , 
\end{align} 
and the symbol ${\boldsymbol D}^{\boldsymbol \eta }$ always denotes (\ref{D^eta}) in the rest. 
(See section 6.2 of \cite{Erler:2015lya}.) 
Then, the Maurer-Cartan element of ${\boldsymbol D}^{\boldsymbol \eta } = {\boldsymbol \eta } - {\boldsymbol m}_{2}|_{0}$ is given by 
\begin{align}  
{\boldsymbol D}^{\boldsymbol \eta } \frac{1}{1-\widetilde{A}} & = \frac{1}{1-\widetilde{A}} \otimes  \pi _{1} \Big( {\boldsymbol D}^{\boldsymbol \eta } \frac{1}{1-\widetilde{A}}  \Big) \otimes \frac{1}{1-\widetilde{A}}
\no \nonumber 
& = \frac{1}{1-\widetilde{A}} \otimes  \pi_{1} \Big( {\boldsymbol \eta } \widetilde{A} - {\boldsymbol m}_{2}|_{0} ( \widetilde{A} , \widetilde{A} ) \Big) \otimes \frac{1}{1-\widetilde{A} } , 
\end{align} 
where $\widetilde{A} = \widetilde{A}^{\rm NS} + \widetilde{A}^{\rm R}$ is a state of the large Hilbert space $\mathcal{H}$ and $\pi _{1}$ is an natural $1$-state projection onto $\mathcal{H}$. 
Hence, the solution of the Maurer-Cartan eqiation ${\boldsymbol D}^{\boldsymbol \eta } (1-\widetilde{A})^{-1} = 0$ is given by a state $\widetilde{A}_{\eta } = \widetilde{A}^{\rm NS}_{\eta } + \widetilde{A}^{\rm R}_{\eta }$ satisfying 
\begin{align} 
\label{EKS pure-gauge-like eq.}
{\boldsymbol \eta } \widetilde{A}_{\eta } - {\boldsymbol m}_{2}|_{0} ( \widetilde{A}_{\eta } , \widetilde{A}_{\eta } ) = 0 , 
\end{align}
and vice versa. 
The solution $\widetilde{A}_{\eta }= \widetilde{A}^{\rm NS}_{\eta } + \widetilde{A}^{\rm R}_{\eta }$ satisfies (\ref{EKS pure-gauge-like eq.}), namely, 
\begin{align} 
\nonumber 
{\rm NS} \, & : \hspace{13mm} \eta \widetilde{A}^{\rm NS}_{\eta } - m_{2} ( \widetilde{A}^{\rm NS}_{\eta } , \widetilde{A}^{\rm NS}_{\eta } ) = 0 ,
\\ \nonumber 
{\rm R} \, & : \hspace{5mm} \eta \widetilde{A}^{\rm R}_{\eta } - m_{2} ( \widetilde{A}^{\rm NS}_{\eta } , \widetilde{A}^{\rm R}_{\eta } ) - m_{2} ( \widetilde{A}^{\rm R}_{\eta } , \widetilde{A}^{\rm NS}_{\eta } ) = 0 , 
\end{align}
which is just equivalent to the condition (\ref{EKS NS pure-gauge-like eq.}) and (\ref{EKS R pure-gauge-like eq.}) characterizing NS and R pure-gauge-like fields $\widetilde{A}^{\rm NS}_{\eta }$ and $\widetilde{A}^{\rm R}_{\eta }$. 
As a result, we obtain one of the most important fact the solutions of the Maurer-Cartan equation of ${\boldsymbol D}^{\boldsymbol \eta } = {\boldsymbol \eta } - {\boldsymbol m}_{2}|_{0}$ gives desired NS and R pure-gauge-like fields.

\niu{NS and R pure-gaugel-like fields} 

When the ${\boldsymbol \eta }$-complex $( \mathcal{H} , {\boldsymbol \eta } )$ is exact, there exist ${\boldsymbol \xi }$ such that $\ld {\boldsymbol \eta } , {\boldsymbol \xi } \rd  = {\boldsymbol 1}$ and $\mathcal{H}$, the large Hilbert space, is decomposed into the direct sum of $\eta $-exacts and $\xi $-exacts $\mathcal{H} = P_{\eta } \mathcal{H} \oplus P_{\xi }\mathcal{H}$, where $P_{\eta }$ and $P_{\xi }$ are projector onto $\eta $-exact and $\xi $-exact states respectively.\footnote{These satisfy $P_{\eta } ^{2} = P_{\eta }$, $P_{\xi }^{2} = P_{\xi }$, $P_{\eta } P_{\xi } = P_{\xi } P_{\eta }=0$,and $P_{\eta } + P_{\xi } = 1$ on $\mathcal{H}$. } 
Note that since the small Hilbert space $\mathcal{H}_{\rm S}$ is defined by $\mathcal{H}_{\rm S} \equiv P_{\eta } \mathcal{H}$ and satisfies $\mathcal{H}_{\rm S} \subset P_{\eta } \mathcal{H}_{\rm S}$, all the states $\widetilde{\Psi }$ belonging to $\mathcal{H}_{\rm S}$ satisfy $P_{\eta } \widetilde{\Psi } = \widetilde{\Psi }$ and $P_{\xi } \widetilde{\Psi } = 0$, or simply, 
\begin{align} 
\nonumber 
{\boldsymbol \eta } \, \widetilde{\Psi } = 0. 
\end{align}
Using this fact, we can construct desired pure-gauge-like (functional) fields $\widetilde{A}^{\rm NS}_{\eta }$ and $\widetilde{A}^{\rm R}_{\eta }$ as solutions of the Maurer-Cartan equation of ${\boldsymbol D}^{{\boldsymbol \eta }} = {\boldsymbol \eta } - {\boldsymbol m}_{2}|_{0}$. 
Note that the Maurer-Cartan equation consists of NS and R out-puts 
\begin{align}
{\rm NS} \, & : \hspace{5mm} 
\pi _{1} {\boldsymbol D}^{\boldsymbol \eta } \frac{1}{1- \widetilde{A}_{\eta }} \Big{|}^{\rm NS} = \pi _{1}  {\boldsymbol D}^{\boldsymbol \eta } \frac{1}{1- \widetilde{A}^{\rm NS}_{\eta }} = 0 , 
\hspace{5mm } 
\no \nonumber 
{\rm R} \, & : \hspace{6.5mm} 
\pi _{1}  {\boldsymbol D}^{\boldsymbol \eta } \frac{1}{1- \widetilde{A}_{\eta }} \Big{|}^{\rm R} = \pi _{1} {\boldsymbol D}^{\boldsymbol \eta } \Big( \frac{1}{1- \widetilde{A}^{\rm NS}_{\eta }} \otimes \widetilde{A}^{\rm R}_{\eta } \otimes \frac{1}{1- \widetilde{A}^{\rm NS}_{\eta }} \Big) = 0 , 
\end{align}
where the upper index of $|$ denotes the NS or R projection: for any state $\widetilde{A} = \widetilde{A}^{\rm NS} + \widetilde{A}^{\rm R} \in \mathcal{H}$, the NS projection $\widetilde{A}|^{\rm NS}$ is defined by $\widetilde{A}|^{\rm NS} \equiv \widetilde{A}^{\rm NS}$ and the R projection $\widetilde{A}|^{\rm R}$ is defined by $\widetilde{A}|^{\rm R} \equiv \widetilde{A}^{\rm R}$.

\vspace{3mm}

An NS pure-gauge-like (functional) field $\widetilde{A}^{\rm NS}_{\eta } = \widetilde{A}^{\rm NS}_{\eta } [ \widetilde{\Psi } ]$ is given by 
\begin{align}
\nonumber 
\widetilde{A}^{\rm NS}_{\eta } \equiv \pi _{1} \widehat{\bf G} \, \frac{1}{1- \widetilde{\Psi }} \Big{|}^{\rm NS} 
= \pi _{1} \widehat{\bf G} \, \frac{1}{1- \widetilde{\Psi }^{\rm NS}}  
\end{align}
because it becomes a trivial NS state solution of the Maurer-Cartan equation as follows 
\begin{align} 
{\boldsymbol D}^{\boldsymbol \eta } \frac{1}{1- \widetilde{A}^{\rm NS}_{\eta }} 
& = {\boldsymbol D}^{\boldsymbol \eta } \frac{1}{1- \pi _{1} \widehat{\bf G} \, \frac{1}{1-\widetilde{\Psi }^{\rm NS} } } = {\boldsymbol D}^{\boldsymbol \eta } \, \widehat{\bf G} \, \frac{1}{1-\widetilde{\Psi }^{\rm NS}} =  \widehat{\bf G} \, {\boldsymbol \eta }  \frac{1}{1-\widetilde{\Psi }^{\rm NS}} 
\no \nonumber 
& = \widehat{\bf G} \, \Big( \frac{1}{1- \widetilde{\Psi }^{\rm NS} } \otimes {\boldsymbol \eta } \widetilde{\Psi }^{\rm NS} \otimes \frac{1}{1-\widetilde{\Psi }^{\rm NS} }  \Big) = 0 . 
\end{align}
Note that $\pi _{1} {\boldsymbol D}^{\boldsymbol \eta } (1- \widetilde{A}^{\rm NS}_{\eta })^{-1} = 0$ is equal to 
\begin{align} 
\nonumber 
\eta \, \widetilde{A}^{\rm NS}_{\eta } - m_{2} ( \widetilde{A}^{\rm NS}_{\eta } , \widetilde{A}^{\rm NS}_{\eta } ) = 0 . 
\end{align}


\vspace{2mm} 

Similarly, a R pure-gauge-like (functional) field $\widetilde{A}^{\rm R}_{\eta }= \widetilde{A}^{\rm R}_{\eta } [ \widetilde{\Psi } ]$ is given by 
\begin{align}
\nonumber 
\widetilde{A}^{\rm R}_{\eta } \equiv \pi _{1} \widehat{\bf G} \, \frac{1}{1- \widetilde{\Psi }} \Big{|}^{\rm R}
= \pi _{1} \widehat{\bf G} \, \Big( \frac{1}{1-\widetilde{\Psi }^{\rm NS}} \otimes \widetilde{\Psi }^{\rm R} \otimes \frac{1}{1-\widetilde{\Psi }^{\rm NS}} \Big)  
\end{align}
because it becomes a trivial R state solution of the Maurer-Cartan equation as follows 
\begin{align} 
 {\boldsymbol D}^{\boldsymbol \eta } \Big( \frac{1}{1- \widetilde{A}^{\rm NS}_{\eta }} \otimes \widetilde{A}^{\rm R}_{\eta } \otimes \frac{1}{1- \widetilde{A}^{\rm NS}_{\eta }} \Big) 
& =  {\boldsymbol D}^{\boldsymbol \eta } \, \widehat{\bf G} \Big( \frac{1}{1- \widetilde{\Psi }^{\rm NS}} \otimes \widetilde{\Psi }^{\rm R} \otimes \frac{1}{1- \widetilde{\Psi }^{\rm NS} } \Big) 
\no \nonumber 
& =  \widehat{\bf G} \, {\boldsymbol \eta } \, \Big( \frac{1}{1- \widetilde{\Psi }^{\rm NS}} \otimes \widetilde{\Psi }^{\rm R} \otimes \frac{1}{1- \widetilde{\Psi }^{\rm NS} } \Big) = 0 . 
\end{align}
Hence, the R state solution $\widetilde{A}^{\rm R}_{\eta }$ satisfies  
\begin{align} 
\nonumber 
\eta \, \widetilde{A}^{\rm R}_{\eta } - \Ld \widetilde{A}^{\rm NS}_{\eta } , \widetilde{A}^{\rm R}_{\eta } \Rd _{\ast } = 0 . 
\end{align}
Note that $\ld \widetilde{A}^{\rm NS}_{\eta } , \widetilde{A}^{\rm R}_{\eta } \rd _{\ast } = m_{2} ( \widetilde{A}^{\rm NS}_{\eta } , \widetilde{A}^{\rm R}_{\eta } ) + m_{2} ( \widetilde{A}^{\rm R}_{\eta } , \widetilde{A}^{\rm NS}_{\eta } )$. 

\vspace{2mm}

\niu{Shift of the dual $A_{\infty }$ products ${\boldsymbol D}^{\boldsymbol \eta }$} 

We introduce the $\widetilde{A}_{\eta }$-shifted products $[ B_{1} , \dots , B_{n} ]^{\eta }_{\widetilde{A}_{\eta }}$ defined by 
\begin{align} 
\big[ B_{1} , \dots , B_{n} \big] ^{\eta }_{\widetilde{A}_{\eta }} \equiv \pi _{1} {\boldsymbol D}^{\boldsymbol \eta } \Big( \frac{1}{1-\widetilde{A}_{\eta } } \otimes B_{1} \otimes \frac{1}{1-\widetilde{A}_{\eta } } \otimes \dots \otimes \frac{1}{1-\widetilde{A}_{\eta } } \otimes B_{n} \otimes \frac{1}{1-\widetilde{A}_{\eta } } \Big) . 
\end{align} 
Note that higher shifted products all vanish $[ B_{1} , \dots , B_{n>2} ]^{\eta }_{\widetilde{A}_{\eta }} =0$ because now we consider ${\boldsymbol D}^{\boldsymbol \eta } \equiv \eta - m_{2}|_{0}$. 
In particular, we write $D_{\eta } B$ for $[ B ]^{\eta }_{A_{\eta }}$: 
\begin{align}
D_{\eta } B \equiv \pi _{1} {\boldsymbol D}^{{\boldsymbol \eta }} \Big( \frac{1}{1-\widetilde{A}_{\eta } } \otimes B \otimes \frac{1}{1-\widetilde{A}_{\eta } } \Big) ,  
\end{align}
or equivalently, for $\widetilde{A}_{\eta } = \widetilde{A}^{\rm NS}_{\eta } + \widetilde{A}^{\rm R}_{\eta }$ and $B = B^{\rm NS} + B^{\rm R}$, 
\begin{align}
\nonumber 
{\rm NS} \, & : \hspace{5mm} D_{\eta } B \big{|}^{\rm NS} = \eta B^{\rm NS} - \Ld \widetilde{A}^{\rm NS}_{\eta } , B^{\rm NS} \Rd _{\ast } , 
\\ \nonumber 
{\rm R} \, & : \hspace{6.5mm} D_{\eta } B \big{|}^{\rm R} = \eta B^{\rm R} - \Ld \widetilde{A}^{\rm NS}_{\eta } , B^{\rm R} \Rd _{\ast } - \Ld \widetilde{A}^{\rm R}_{\eta } , B^{\rm NS} \Rd _{\ast } . 
\end{align}
When $\widetilde{A}_{\eta }$ gives a solution on the Maurer-Cartan equation of ${\boldsymbol D}_{\boldsymbol \eta}$, these $\widetilde{A}_{\eta }$-shifted products also satisfy $A_{\infty }$-relations, which implies that the linear operator $D_{\eta }$ becomes nilpotent. 
We find 
\begin{align}
( D_{\eta } )^{2 } B & = \pi _{1} {\boldsymbol D}^{\boldsymbol \eta } \Big( \frac{1}{1-\widetilde{A}_{\eta } }  \otimes \pi _{1} {\boldsymbol D}^{\boldsymbol \eta } \Big( \frac{1}{1-\widetilde{A}_{\eta } } \otimes B \otimes \frac{1}{1-\widetilde{A}_{\eta } } \Big) \otimes \frac{1}{1-\widetilde{A}_{\eta } } \Big) 
\no \nonumber 
& = \pi _{1} ( {\boldsymbol D}^{\boldsymbol \eta } )^{2} \frac{1}{1-\widetilde{A}_{\eta } } - \LD \pi _{1} {\boldsymbol D}^{\boldsymbol \eta } \Big( \frac{1}{1-\widetilde{A}_{\eta } } \Big) , \,  B \, \RD ^{\eta }_{\ast \widetilde{A}_{\eta }}  = 0 .  
\end{align}

\vspace{2mm}

\niu{NS and R associated fields}

Let ${\boldsymbol d}$ be a coderivation constructed from a derivation $d$ of the dual $A_{\infty }$-products ${\boldsymbol D}^{\boldsymbol \eta }$, which implies that the $d$-derivation propery $\ld {\boldsymbol d} , {\boldsymbol D}^{\boldsymbol \eta } \rd = 0$ holds. 
Then, we obtain $\ld {\boldsymbol D}_{\boldsymbol d} , {\boldsymbol \eta } \rd = 0 $ with ${\boldsymbol D}_{\boldsymbol d} \equiv \widehat{\bf G}^{-1} \, {\boldsymbol d} \, \widehat{\bf G}$, which means that ${\boldsymbol D}_{\boldsymbol d}$ is ``${\boldsymbol \eta }$-exact'' and there exists a coderivation ${\boldsymbol \xi }_{\boldsymbol d}$ such that 
\begin{align}
{\boldsymbol D}_{\boldsymbol d}= \widehat{\bf G}^{-1} \, {\boldsymbol d} \, \widehat{\bf G} = ( - )^{{\boldsymbol d}} \ld {\boldsymbol \eta } , {\boldsymbol \xi }_{\boldsymbol d} \rd . 
\end{align} 
Using this coderivation ${\boldsymbol \xi}_{\boldsymbol d}$, we can construct NS and R associated fields constructed from the derivation operator $d$. 
Note that the response of ${\boldsymbol d}$ acting on the group-like element of $\widetilde{A}_{\eta }=\widetilde{A}^{\rm NS}_{\eta } + \widetilde{A}^{\rm R}_{\eta }$ is given by 
\begin{align}
(-)^{{\boldsymbol d}} {\boldsymbol d} \, \frac{1}{1-\widetilde{A}_{\eta } } & = (-)^{{\boldsymbol d}} G G^{-1} \, {\boldsymbol d} \, G   \frac{1}{1- \widetilde{\Psi } } =  G \, {\boldsymbol \eta } \, {\boldsymbol \xi }_{{\boldsymbol d}} \frac{1}{1- \widetilde{\Phi } } = {\boldsymbol D}^{{\boldsymbol \eta }} \, G \Big( {\boldsymbol \xi }_{{\boldsymbol d}} \frac{1}{1- \widetilde{\Psi } } \Big) 
\no 
&= {\boldsymbol D}^{{\boldsymbol \eta }} \Big( \frac{1}{1-\widetilde{A}_{\eta } } \otimes  \pi _{1} G \Big( {\boldsymbol \xi }_{{\boldsymbol d}} \frac{1}{1- \widetilde{\Psi }} \Big)  \otimes \frac{1}{1-\widetilde{A}_{\eta }} \Big) . 
\end{align}

\vspace{2mm} 

An NS associated (functional) field $\widetilde{A}^{\rm NS}_{d} = \widetilde{A}^{\rm NS}_{d} [ \widetilde{\Psi } ]$ of $d$ is given by 
\begin{align}
\nonumber 
\widetilde{A}^{\rm NS}_{d} & \equiv \pi _{1} G \Big( {\boldsymbol \xi }_{{\boldsymbol d}} \frac{1}{1- \widetilde{\Psi } } \Big) \Big{|}^{\rm NS} 
\end{align}
because one can directly check 
\begin{align} 
(-)^{{\boldsymbol d}} {\boldsymbol d} \, \frac{1}{1-\widetilde{A}^{\rm NS}_{\eta } } & = (-)^{{\boldsymbol d}} {\boldsymbol d} \, \frac{1}{1-\widetilde{A}_{\eta } } \Big{|}^{\rm NS} 
= {\boldsymbol D}^{{\boldsymbol \eta }} \Big( \frac{1}{1-\widetilde{A}_{\eta } } \otimes  \pi _{1} G \Big( {\boldsymbol \xi }_{{\boldsymbol d}} \frac{1}{1- \widetilde{\Psi }} \Big)  \otimes \frac{1}{1-\widetilde{A}_{\eta }} \Big) \Big|^{\rm NS} 
\no \nonumber 
& = {\boldsymbol D}^{{\boldsymbol \eta }} \Big( \frac{1}{1-\widetilde{A}^{\rm NS}_{\eta } } \otimes  \pi _{1} G \Big( {\boldsymbol \xi }_{{\boldsymbol d}} \frac{1}{1- \widetilde{\Psi }} \Big) \Big{|}^{\rm NS}  \otimes \frac{1}{1-\widetilde{A}^{\rm NS}_{\eta }} \Big) . 
\end{align}
Picking up the relation on $\mathcal{H}$, or equivalently acting $\pi _{1}$ on this relation on $T(\mathcal{H})$, we obtain  
\begin{align} 
\nonumber 
(-)^{d} d  \, \widetilde{A}^{\rm NS}_{\eta } & =  \eta \, \widetilde{A}^{\rm NS}_{d}  - \Ld \widetilde{A}^{\rm NS}_{\eta } , \widetilde{A}^{\rm NS}_{d} \Rd _{\ast } ,
\end{align}
which is the simplest case of $(-)^{d} d \widetilde{A}_{\eta } = \pi _{1} {\boldsymbol D}^{\boldsymbol \eta } \mathchar`- {\rm exact} \,\, {\rm term}$. 

\vspace{2mm} 

Similarly, an R associated (functional) field $\widetilde{A}^{\rm R}_{d} = \widetilde{A}^{\rm R}_{d} [ \widetilde{\Psi } ]$ of $d$ is given by 
\begin{align}
\nonumber 
\widetilde{A}^{\rm R}_{d} & \equiv \pi _{1} G \Big( {\boldsymbol \xi }_{{\boldsymbol d}} \frac{1}{1- \widetilde{\Psi } } \Big) \Big{|}^{\rm R} 
\end{align}
because one can directly check 
\begin{align} 
(-)^{d} d \, \widetilde{A}^{\rm R}_{\eta } & = \pi _{1} (-)^{{\boldsymbol d}} {\boldsymbol d} \, \frac{1}{1-\widetilde{A}_{\eta } } \Big{|}^{\rm R} 
= \pi _{1} {\boldsymbol D}^{{\boldsymbol \eta }} \Big( \frac{1}{1-\widetilde{A}_{\eta } } \otimes  \pi _{1} G \Big( {\boldsymbol \xi }_{{\boldsymbol d}} \frac{1}{1- \widetilde{\Psi }} \Big)  \otimes \frac{1}{1-\widetilde{A}_{\eta }} \Big) \Big|^{\rm R} 
\no 
& = \eta \, \pi _{1} G \Big( {\boldsymbol \xi }_{{\boldsymbol d}} \frac{1}{1- \widetilde{\Psi } } \Big) \Big{|}^{\rm R}  - \LD \widetilde{A}^{\rm NS}_{\eta } , \pi _{1} G \Big( {\boldsymbol \xi }_{{\boldsymbol d}} \frac{1}{1- \widetilde{\Psi } } \Big) \Big{|}^{\rm R} \RD _{\ast } - \LD \widetilde{A}^{\rm R}_{\eta } , \pi _{1} G \Big( {\boldsymbol \xi }_{{\boldsymbol d}} \frac{1}{1- \widetilde{\Psi } } \Big) \Big{|}^{\rm NS} \RD _{\ast } 
\no \nonumber 
& = \eta \widetilde{A}^{\rm R}_{d} - \Ld \widetilde{A}^{\rm NS}_{\eta } , \widetilde{A}^{\rm R}_{d} \Rd _{\ast } - \Ld \widetilde{A}^{\rm R}_{\eta } , \widetilde{A}^{\rm NS}_{d} \Rd _{\ast } . 
\end{align} 
We obtain $(-)^{d} d \widetilde{A}^{\rm R}_{\eta } = D^{\rm NS}_{\eta } \widetilde{A}^{\rm R}_{d} + \ld \widetilde{A}^{\rm R}_{\eta } , \widetilde{A}^{\rm NS}_{d} \rd _{\ast } $, namely, $(-)^{d} d \widetilde{A}_{\eta } = \pi {\boldsymbol D}^{\boldsymbol \eta } \mathchar`- {\rm exact} \,\, {\rm terms}$. 

\vspace{2mm} 

We checked that $\widetilde{A}^{\rm NS}_{\eta }$, $\widetilde{A}^{\rm R}_{\eta }$ defined by (\ref{NS pure-gauge-like tilde}), (\ref{R pure-gauge-like tilde}) and $\widetilde{A}^{\rm NS}_{d}$, $\widetilde{A}^{\rm R}_{d}$ defined by (\ref{NS associated tilde}), (\ref{R associated tilde}) indeed satisfy WZW-like relations (\ref{EKS NS pure-gauge-like eq.}), (\ref{EKS R pure-gauge-like eq.}), (\ref{NS associated eq. tilde}), and (\ref{R associated eq. tilde}). 
Namely, (\ref{S tilde}) is consistent. 

\vspace{2mm}

\subsection{Equivalence of EKS and KO theories} 

In section 2, for given dynamical string field $\varphi $, using functional fields $\mathcal{A}_{\eta } [\varphi ]$ parametrized by $\varphi $, we proposed the Wess-Zumino-Witten-like complete action 
\begin{align} 
\nonumber 
S_{\rm wzw} [\varphi ] = - \int_{0}^{1} dt \, \langle \mathcal{A}^{\ast }_{t} , Q \mathcal{A}_{\eta } + m_{2}|_{2} ( \mathcal{A}_{\eta } , \mathcal{A}_{\eta } ) \rangle , 
\end{align} 
where ${\cal A}_{\eta } \equiv {\cal A}^{\rm NS}_{\eta } + {\cal A}^{\rm R}_{\eta }$ and ${\cal A}^{\ast }_{t} \equiv {\cal A}^{\rm NS}_{t} + \xi Y \partial _{t} ( P_{\eta } {\cal A}^{\rm R}_{\eta } ) = (F \xi + Y \xi )\partial _{t} \mathcal{A}_{\eta }$. 

\vspace{2mm} 

We found that one realization of this WZW-like complete action is given by setting 
\begin{align} 
\label{KO}
\mathcal{A}^{\rm NS}_{\eta } [\Phi ^{\rm NS} ] : =  \big( \eta e^{\Phi ^{\rm NS}} \big) e^{- \Phi ^{\rm NS} } \equiv A^{\rm NS}_{\eta } , \hspace{8mm} \mathcal{A}^{\rm R}_{\eta } [\Phi ^{\rm NS} , \Psi ^{\rm R} ] : =   F \Psi ^{\rm R} \equiv  A^{\rm R}_{\eta } , 
\end{align}
which is just Kunitomo-Okawa's action proposed in \cite{Kunitomo:2015usa}. 
This is the WZW-like theory $S_{\rm wzw} = S_{\rm wzw} [\Phi ^{\rm NS}, \Psi ^{\rm R} ]$ parametrized by $\varphi |^{\rm NS} = \Phi ^{\rm NS}$, $\varphi |^{\rm R} = \Psi ^{\rm R}$. 
Another realization of the action, which was proposed in section 3.1 and checked in section 3.2, is given by setting 
\begin{align}
\label{EKS}
\mathcal{A}^{\rm NS}_{\eta } [ \widetilde{\Psi } ] : = \pi _{1} \widehat{\bf G} \frac{1}{1- \widetilde{\Psi } }\Big{|}^{\rm NS} \equiv \widetilde{A}^{\rm NS}_{\eta } , \hspace{5mm} 
\mathcal{A}^{\rm R}_{\eta } [\widetilde{\Psi }] : =  \pi _{1} \widehat{\bf G} \frac{1}{1- \widetilde{\Psi } }\Big{|}^{\rm R} \equiv \widetilde{A}^{\rm R}_{\eta } , 
\end{align} 
which reproduces the Ramond equations of motion proposed in \cite{Erler:2015lya}. 
This is the WZW-like theory $S_{\rm wzw} = S_{\rm wzw} [\widetilde{\Psi } ]$ parametrized by $\varphi = \widetilde{\Psi }$. 
Note also that the kinetic terms of (\ref{S KO}) and $(\ref{S tilde})$ have the same spectrum. 
As a result, we obtain the equivalence of two theories proposed in \cite{Kunitomo:2015usa} and \cite{Erler:2015lya}, which are different parametrizations of (\ref{WZW-like complete action}). 
See also \cite{Erler:2015rra, GM2}. 

\vspace{2mm} 

In other words, since both (\ref{KO}) and (\ref{EKS}) have the same WZW-like structure and gives the same WZW-like action (\ref{WZW-like complete action}), we can identify $A_{\eta } = A^{\rm NS}_{\eta } + A^{\rm R}_{\eta }$ and $\widetilde{A}_{\eta }= \widetilde{A}^{\rm NS}_{\eta } + \widetilde{A}^{\rm R}_{\eta }$ in the same way as \cite{Erler:2015rra}. 
Then, the identification of pure-gauge-fields 
\begin{align} 
\label{parametrizations}
A_{\eta } = \big( \eta e^{\Phi ^{\rm NS} } \big) e^{- \Phi ^{\rm NS} } + F \Psi ^{\rm R}  \cong  \pi _{1} \widehat{\bf G} \frac{1}{1- \widetilde{\Psi } }\Big{|}^{\rm NS} + \pi _{1} \widehat{\bf G} \frac{1}{1- \widetilde{\Psi } }\Big{|}^{\rm R} = \widetilde{A}_{\eta }  
\end{align}
trivially provides the equivalence of two actions (\ref{S KO}) and (\ref{S tilde}): 
The single functional form of (\ref{Single functional form}) gives the equivalence of two theories. 
When we use the WZW-like form of (\ref{WZW-like complete action}), it seems that (\ref{parametrizations}) indirectly gives the equivalence and does not {\it directly} give a field redefinition between two theories. 
A partial gauge fixing $\Phi ^{\rm NS} = \xi \Psi ^{\rm NS}$ is necessitated to directly relate $(\Psi ^{\rm NS} , \Psi ^{\rm R})$ and $(\widetilde{\Psi }^{\rm NS} , \widetilde{\Psi }^{\rm R})$. 
See also \cite{Erler:2015uba, Erler:2015uoa, Goto:2015hpa, GM2}. 
Similarly, as demonstrated in \cite{Erler:2015uoa}, if we start with 
\begin{align}
\label{large redef}
 A_{t} & = \big( \partial _{t} e^{\Phi ^{\rm NS}} \big) e^{- \Phi ^{\rm NS}} + F \Xi \Big( \partial _{t} \Psi ^{\rm R} - \Ld  \big( \partial _{t} e^{\Phi ^{\rm NS}} \big) e^{-\Phi ^{\rm NS}} , F \Psi ^{\rm R} \Rd _{\ast } \Big) 
\no 
& \cong \pi _{1} \widehat{\bf G} \Big( {\boldsymbol \xi}_{t} \frac{1}{1-\widetilde{\Psi }} \Big) \Big{|}^{\rm NS} +  \pi _{1} \widehat{\bf G} \Big( {\boldsymbol \xi}_{t} \frac{1}{1-\widetilde{\Psi }} \Big) \Big{|}^{\rm R}  = \widetilde{A}_{t} , 
\end{align}
we can quickly read a field redefinition of $(\Phi ^{\rm NS} , \Psi ^{\rm R})$ and $(\widetilde{\Phi }^{\rm NS} , \widetilde{\Psi }^{\rm R})$ in the large Hilbert space for NS fields and in the restricted small space for R fields with a trivial up-lift $\widetilde{\Psi }^{\rm NS} = \eta \widetilde{\Phi }^{\rm NS}$. 
One can check that the same logic used for the NS sector in \cite{Erler:2015uoa} also goes in the case including the R sector because WZW-like relations exist as we explained, which is in appendix A. 

\section{Conclusion} 

In this paper, we have clarified a Wess-Zumino-Witten-like structure including Ramond fields and proposed one systematic way to construct gauge invariant actions, which we call WZW-like complete action. 
In this framework, once a WZW-like functional $\mathcal{A}_{\eta } = \mathcal{A}_{\eta } [\varphi ]$ of some dynamical string field $\varphi $ is constructed, one obtain one realization of our WZW-like complete action $S_{\rm wzw}[\varphi ]$ parametrized by $\varphi $. 
On the basis of this way, we have constructed an action $\widetilde{S}$ whose on-shell condition is equivalent to the Ramond equations of motion proposed in \cite{Erler:2015lya}. 
In particular, this action $\widetilde{S} = S_{\rm wzw}[\widetilde{\Psi }]$ and Kunitomo-Okawa's action $S_{\rm KO} = S_{\rm wzw} [\Phi ^{\rm NS}, \Psi ^{\rm R}]$ proposed in \cite{Kunitomo:2015usa} just give different parametrizations of the same WZW-like structure and action, which implies the equivalence of two theories \cite{Kunitomo:2015usa, Erler:2015lya}. 
Let us conclude by discussing future directions. 

\niu{Closed superstring field theories} 

It would be interesting to extend the result of \cite{Kunitomo:2015usa} to closed superstring field theories\cite{GKMO}. 
We expect that our idea of WZW-like structure and action also goes in heterotic and type II theories if the kinetic terms are given by the same form. 
Then, we need explicit expressions of ${\boldsymbol D}^{\boldsymbol \eta }$ and ${\boldsymbol l}$, where ${\boldsymbol D}^{\boldsymbol \eta }$ is a dual $L_{\infty }$ structure of the original $L_{\infty }$ products $\widetilde{\bf L}=\widehat{\bf G}^{-1} \, {\boldsymbol l} \, \widehat{\bf G}$ given in \cite{Erler:2015lya}. 
NS and NS-NS parts of these dual $A_{\infty }$/$L_{\infty }$ structures are discussed in \cite{GM2}.

\niu{Quantization and Supermoduli} 

We would have to quantize the (WZW-like) complete action and clarify its relation with supermoduli of super-Riemann surfaces\cite{Verlinde:1987sd, D'Hoker:1988ta, Saroja:1992vw, Belopolsky:1997jz, Witten:2012bh} to obtain a better understandings of superstrings from recent developments in field theoretical approach. 
The Batalin-Vilkovisky formalism\cite{Batalin:1981jr, Batalin:1984jr} is one helpful way to tackle these problems: A quantum master action is necessitated. 
As a first step, it is important to clarify whether we can obtain an $A_{\infty }$-morphism $\widehat{\bf G}$ which has the cyclic property consistent with the $XY$-projection. 
If it is possible, the resultant action would have an $A_{\infty }$ form and then the classical Batalin-Vilkovisky quantization is straightforward. 
A positive answer is now provided in
\cite{EOT2} for open superstring field theory without stubs. 
It would also be helpful to clarify more detailed relations between recent important developments. 

\section*{Acknowledgement} 

The author would like to thank Keiyu Goto, Hiroshi Kunitomo, and Yuji Okawa for comments. 
The author also would like to express his gratitude to his doctors, nurses, and all the staffs of University Hospital, Kyoto Prefectural University of Medicine, for medical treatments and care during his long hospitalization. 
This work was supported in part by Research Fellowships of the Japan Society for the Promotion of Science for Young Scientists.

\appendix

\section{Basic facts and some identities}

We summarize important properties of the BPZ inner product and give proofs of some relations which we skipped in the text.

\niu{BPZ properties}

The BPZ inner product $\langle A , B \rangle $ in the large Hilbert space of any $A,B \in \mathcal{H}$ has the following properties with the BRST operator $Q$ and the Witten's star product $m_{2}$: 
\begin{align}
\langle A , B \rangle & = (-)^{AB} \langle B , A \rangle , 
\no 
\langle A , Q B \rangle & = - (-)^{A} \langle Q A , B \rangle , 
\no \nonumber  
\langle A , m_{2} (  B , C ) \rangle & = (-)^{A(B+C)} \langle B , m_{2} ( C ,A ) \rangle . 
\end{align} 
Note also that with a projector onto $\eta $-exact states $P_{\eta }$ and $P_{\xi } = 1 - P_{\eta }$ and the zero mode of $\eta \equiv \eta _{0}$ of $\eta (z)$-current, the BPZ inner product satisfies 
\begin{align}
\nonumber 
\langle P_{\eta} A , B \rangle &= \langle A , P_{\xi } B \rangle , 
\no \nonumber 
\langle A , \eta B \rangle &= (-)^{A} \langle \eta A , B \rangle . 
\end{align}
Similarly, for any states in the restricted small space $A^{\rm R}, B^{\rm R} \in \mathcal{H}_{R}$, the bilinear $\langle \xi Y A^{\rm R} ,B^{\rm R} \rangle $ has the following properties: 
\begin{align}
\langle \xi Y A^{\rm R} , Q B^{\rm R} \rangle & = (-)^{AB} \langle \xi Y B^{\rm R} , A^{\rm R} \rangle , 
\no \nonumber 
\langle \xi Y A ^{\rm R} , Q B^{\rm R} \rangle & = - (-)^{A} \langle \xi Y Q A^{\rm R} , B^{\rm R} \rangle . 
\end{align}

\niu{Associated fields in Kunitomo-Okawa theory}

In the work of \cite{Kunitomo:2015usa}, for any state $B\in \mathcal{H}$, the linear map $F$ is defined by
\begin{align}
\nonumber 
F B \equiv \sum_{n=0}^{\infty } \big(  \Xi \Ld A^{\rm NS}_{\eta } , \hspace{3mm} \Rd _{\ast } \big) ^{n} B = \frac{1}{1 - \Xi (\eta - D^{\rm NS}_{\eta } )} B , 
\end{align} 
where $D^{\rm NS}_{\eta } B = \eta B - \ld A^{\rm NS}_{\eta } , B \rd _{\ast } $ and $(D^{\rm NS}_{\eta })^{2} = 0$. 
Thus, its inverse is given by $F^{-1} = \eta \Xi + \Xi D^{\rm NS}_{\eta }$, which provides $\eta F^{-1} = F^{-1} D^{\rm NS}_{\eta }$ or equivalently, $F \eta F^{-1} = D^{\rm NS}_{\eta }$, and thus $\ld D^{\rm NS}_{\eta } , F \Xi \rd _{\ast } = 1$. 
Then, we find that $A^{\rm R}_{\eta } \equiv F \Psi ^{\rm R} $ satisfies $D^{\rm NS}_{\eta } A^{\rm R}_{\eta } = 0$ as follows: 
\begin{align}
\nonumber 
A^{\rm R}_{\eta } \equiv F \Psi ^{\rm R} = F \eta \xi \Psi ^{\rm R} = D^{\rm NS}_{\eta } F \Xi \Psi ^{\rm R} = D^{\rm NS}_{\eta } F \Xi A^{\rm R}_{\eta } . 
\end{align}
With $F^{-1} = \eta \Xi + \Xi D^{\rm NS}_{\eta }$ and $\ld d , \eta \rd = 0$, a derivation $d$ acts on the state $F\Psi $ as
\begin{align} 
d ( F \Psi ) & = F ( d \Psi ) - F \ld d , F^{-1} \rd F \Psi 
\no 
& = F ( d \Psi ) - F \ld d , \eta \Xi \rd F \Psi - F \ld d , \Xi D^{\rm NS}_{\eta } \rd F \Psi  
\no  
& = F ( d \Psi ) - (-)^{d} F \big( \eta \ld d , \Xi \rd + (-)^{d} \ld d , \Xi \rd D^{\rm NS}_{\eta } \big) F \Psi +  F \Xi D^{\rm NS}_{\eta } \ld  A^{\rm NS}_{d} , F \Psi \rd _{\ast }   
\no \nonumber 
& = D^{\rm NS}_{\eta } F \Xi \big( d \Psi - \ld  A^{\rm NS}_{d} , F \Psi \rd _{\ast } - (-)^{d} \eta \ld d , \Xi \rd F \Psi \big) +  \ld  A^{\rm NS}_{d} , F \Psi \rd _{\ast } , 
\end{align}
where $\Psi $ is an arbitrary state. 
We thus obtain 
\begin{align}
\nonumber 
(-)^{d} d A^{\rm R}_{\eta } = D^{\rm NS}_{\eta } A^{\rm R}_{d} - \ld A^{\rm R}_{\eta } , A^{\rm NS}_{d} \rd _{\ast } 
\end{align}
where $A^{\rm R}_{\eta } \equiv F \Psi ^{\rm R} $ is an R pure-gauge-like field and an R associated field $A^{\rm R}_{d}$ is defined by 
\begin{align} 
\nonumber 
A^{\rm R}_{d} & \equiv F \Xi \Big( (-)^{d} d \Psi ^{\rm R} + \ld A^{\rm R}_{\eta } , A^{\rm NS}_{d} \rd _{\ast } + \eta \ld d , \Xi \rd A^{\rm R}_{\eta } \Big) .  
\end{align}

\niu{The original form of Kunitomo-Okawa's action} 

The original form of Kunitomo-Okawa's complete action is 
\begin{align}
\nonumber 
S [\Phi ^{\rm NS} , \Psi ^{\rm R} ] = - \frac{1}{2} \langle \xi Y \Psi ^{\rm R} , Q \Psi ^{\rm R}  \rangle - \int_{0}^{1} dt \, \langle  A^{\rm NS}_{t} , Q A^{\rm NS}_{\eta } + m_{2} ( A^{\rm R}_{\eta } , A^{\rm R}_{\eta } ) \rangle , 
\end{align} 
where $A^{\rm R}_{\eta } = F(t) \Psi ^{\rm R}$. 
Note that $F(t)$ satisfies $F(t=0)=0$ and $F(t=1) = F$. 
Using the cyclic property of the star product $m_{2}$, $A^{\rm R}_{\eta } = D^{\rm NS}_{\eta } F(t) \Xi A^{\rm R}_{\eta }$, and $\ld D^{\rm NS}_{\eta } , F \Xi \rd =1$, we find 
\begin{align}
- \int_{0}^{1} dt \, \langle A^{\rm NS}_{t} , m_{2} ( A^{\rm R}_{\eta } , A^{\rm R}_{\eta } )  \rangle 
& = \frac{1}{2} \int_{0}^{1} dt \,  \langle A^{\rm R}_{\eta }  ,  \ld A^{\rm NS}_{t} , A^{\rm R}_{\eta } \rd _{\ast }  \rangle  
= \frac{1}{2} \int_{0}^{1} dt \,  \langle D^{\rm NS}_{\eta } F (t) \Xi \Psi ,  \ld A^{\rm NS}_{t} , A^{\rm R}_{\eta } \rd _{\ast }  \rangle  
\no 
& = \frac{1}{2} \int_{0}^{1} dt \,  \langle \Psi ^{\rm R} , F (t) \Xi  D^{\rm NS}_{\eta } \ld A^{\rm NS}_{t} , A^{\rm R}_{\eta } \rd _{\ast }  \rangle  
= \frac{1}{2} \int_{0}^{1} dt \,  \langle \Psi ^{\rm R} ,  \partial _{t} ( F (t) \Psi ^{\rm R} ) \rangle  
\no \nonumber 
& = \frac{1}{2} \langle \Psi ^{\rm R} , F \Psi ^{\rm R} \rangle 
= - \frac{1}{2} \langle \xi Y \Psi ^{\rm R} , \eta X F \Psi ^{\rm R} \rangle . 
\end{align}
Note also that $XY \Psi ^{\rm R}= \Psi ^{\rm R}$ and $\eta \xi \Psi ^{\rm R} = \Psi ^{\rm R}$. 
Hence, we find that 
\begin{align}
\nonumber 
- \frac{1}{2} \langle \xi Y \Psi ^{\rm R} , Q \Psi ^{\rm R}  \rangle - \int_{0}^{1} dt \, \langle A^{\rm NS}_{t} , m _{2} ( F(t) \Psi ^{\rm R} , F(t) \Psi ^{\rm R} ) \rangle 
& = - \frac{1}{2} \langle \xi Y \Psi ^{\rm R} , \, Q \Psi ^{\rm R} + \eta X F \Psi ^{\rm R} \rangle 
\no 
& = - \frac{1}{2} \langle \xi Y \Psi ^{\rm R} , Q F \Psi ^{\rm R} \rangle . 
\end{align}
As we explained in section 1.1, this is equal to (\ref{S KO}). 

\niu{Identification of $A_{t} \cong \widetilde{A}_{t}$ provides $A_{\eta } = \widetilde{A}_{\eta }$}

We check that the identification $A_{t} = A^{\rm NS}_{t}+ A^{\rm R}_{t}$ and $\widetilde{A}_{t}=\widetilde{A}^{\rm NS}_{t} + \widetilde{A}^{\rm R}_{t}$ provide a field redefinition of $(\Phi ^{\rm NS} , \Psi ^{\rm R})$ and $(\widetilde{\Phi }^{\rm NS} , \widetilde{\Psi }^{\rm R})$ with $\widetilde{\Psi }^{\rm NS} = \eta \widetilde{\Phi }^{\rm NS}$.   
We start is $A_{t} - \widetilde{A}_{t} = 0$. 
Then the relation $\eta ( A_{t} - \widetilde{A}_{t} ) = 0$ automatically holds. 
Recall that we have WZW-like relation $\partial _{t} A_{\eta } = D_{\eta } A_{t}$ and $\partial _{t} \widetilde{A}_{\eta } = \widetilde{D}_{\eta } A_{t}$ where $D_{\eta } B = \eta B - m_{2}|_{0} ( A_{\eta } , B) - (-)^{A_{\eta } B} m_{2}|_{0} ( B , A_{\eta })$. 
Therefore, using these WZW-like relations and the identification $A_{t} = \widetilde{A}_{t}$, one can rewrite $\eta (A_{t} - \widetilde{A}_{t})=0$ as 
\begin{align}
\label{Eq}
\partial _{t} (A_{\eta } - \widetilde{A}_{\eta } ) = m_{2}|_{0} \big( A_{\eta } -\widetilde{A}_{\eta } , A_{t} \big) - m_{2}|_{0} \big( A_{t} , A_{\eta } -\widetilde{A} \big) . 
\end{align}
For brevity, we define ${\cal I}^{\rm NS} (t) \equiv A^{\rm NS}_{\eta } (t) - \widetilde{A}^{\rm NS}_{\eta } (t)$ and ${\cal I}^{\rm R}  (t) \equiv A^{\rm R}_{\eta } (t) - \widetilde{A}^{\rm R}_{\eta } (t)$. 
Note that the $t=0$ values $A_{\eta } (t=0) = \widetilde{A}_{\eta } (t=0) = 0$ gives the initial conditions $\mathcal{I}^{\rm NS} (t=0)=0$ and $\mathcal{I}^{\rm R}(t=0)=0$. 
Then, the NS output state and R output state of (\ref{Eq}) can be separated as 
\begin{align}
\label{Eq NS}
{\rm NS} \, : & \hspace{15mm} \frac{\partial }{\partial t} \, \mathcal{I}^{\rm NS} (t) =  \Ld \mathcal{I}^{\rm NS} (t) , \, A^{\rm NS}_{t} (t) \Rd _{\ast } , 
\\ \label{Eq R}
{\rm R} \, : & \hspace{5mm} \frac{\partial }{\partial t} \, \mathcal{I}^{\rm R} (t) =  \Ld \mathcal{I}^{\rm R} (t), \, A^{\rm NS}_{t} (t) \Rd _{\ast } + \Ld \mathcal{I}^{\rm NS} (t) , \, A^{\rm R}_{t} (t) \Rd _{\ast } . 
\end{align} 
The initial condition $\mathcal{I}^{\rm NS}(t=0) = 0$ provides the solution $\mathcal{I}^{\rm NS} (t) = 0$ of the differential equation (\ref{Eq NS}), which means $A_{\eta }^{\rm NS} = \widetilde{A}^{\rm NS}_{\eta }$. 
Then, the R output equation (\ref{Eq R}) reduces to 
\begin{align}
\nonumber 
{\rm R} \, : & \hspace{5mm} \frac{\partial }{\partial t} \, \mathcal{I}^{\rm R} (t) =  \Ld \mathcal{I}^{\rm R} (t), \, A^{\rm NS}_{t} (t) \Rd  _{\ast } 
\end{align} 
and the initial condition $\mathcal{I}^{\rm R}(t = 0) = 0$ also provides the solution $\mathcal{I}^{\rm R} (t) = 0$, which implies $A^{\rm R}_{\eta } = \widetilde{A}^{\rm R}_{\eta }$. 
As a result, under the identification $A_{t} \cong \widetilde{A}_{t}$, we obtain $A_{\eta } = \widetilde{A}_{\eta }$. 

\niu{Field relation of $(\Phi ^{\rm NS} , \Psi ^{\rm R})$ and $(\widetilde{\Phi }^{\rm NS} , \widetilde{\Psi }^{\rm R} )$}

Under the identification $A_{t} \cong \widetilde{A}_{t}$, we obtained $A_{\eta } = \widetilde{A}_{\eta}$, which provides 
\begin{align} 
\label{Rel}
\frac{1}{1 - A_{\eta }} \otimes A_{t} \otimes \frac{1}{1-A_{\eta }} = \widehat{\bf G} \frac{1}{1-\widetilde{\Psi } } \otimes  {\boldsymbol \xi }_{t} \widetilde{\Psi } \otimes \frac{1}{1-\widetilde{\Psi }} , 
\end{align}
where $\widetilde{\Psi } = \eta \widetilde{\Phi }^{\rm NS} + \widetilde{\Psi }^{\rm R}$ and ${\boldsymbol \xi}_{t} = \xi \partial _{t}$. 
One can read the NS and R outputs of (\ref{Rel}) as 
\begin{align}
{\rm NS} \, : & \hspace{5mm} \widetilde{\Phi }^{\rm NS} = \pi _{1} \, \int_{0}^{1} dt \, \Big( \widehat{\bf G}^{-1} \frac{1}{1 - A_{\eta }(t)} \otimes A_{t} (t) \otimes \frac{1}{1-A_{\eta } (t) }  \Big) \Big{|}^{\rm NS} , 
\\ 
{\rm R} \, : & \hspace{5mm} \widetilde{\Psi }^{\rm R} = \pi _{1} \, {\boldsymbol \eta } \, \int_{0}^{1} dt \, \Big( \widehat{\bf G}^{-1} \frac{1}{1 - A_{\eta }(t) } \otimes A_{t} (t)  \otimes \frac{1}{1-A_{\eta }(t)}  \Big) \Big{|}^{\rm R} . 
\end{align}



\end{document}